\documentclass[
  aps,
  pre,
  reprint,
  nofootinbib,
  longbibliography
]{revtex4-2}

\usepackage{amsmath,amssymb,mathtools}
\usepackage{graphicx}
\usepackage{microtype}
\usepackage[
    colorlinks=true,
    allcolors=blue
]{hyperref}

\newcommand{\E}{\mathbb{E}}
\newcommand{\Pp}{\mathbb{P}}
\newcommand{\ind}{\mathbf{1}}

\newcommand{\Exp}{\mathrm{Exp}}
\newcommand{\dd}{\mathrm{d}}

\begin{document}

\title{Exact mean first-passage time and a Poisson-smoothing crossover in renewal shot noise}
\author{Yuxin Wang}
\author{Jiajun Zhang}
\affiliation{School of Mathematics, Sun Yat-sen University, Guangzhou, China}

\date{\today}

\begin{abstract}
Threshold crossings in exponentially relaxing shot noise depend on the interplay between incoming impulses and the residual signal left by earlier arrivals. For renewal input, existing high-threshold product approximations of the mean first-passage time (MFPT) leave open when evaluation at a single index preserves the mean with vanishing relative error. Here we derive an exact finite-threshold MFPT series for exponentially marked renewal shot noise and recast it as an average over an auxiliary Poisson index. When the waiting-time density exhibits short-time power-law behavior, $w(t)\sim ct^{\kappa-1}$, the dominant series contributions shift from the Poisson center by an amount of order $x^{1-\kappa}$, whereas the averaging kernel has a fluctuation width of order $\sqrt{x}$, with $x=\lambda b$. The competition between these scales yields a crossover at $\kappa=1/2$, where the deterministic-index product is asymptotically exact in ratio above this boundary, requires a finite multiplicative correction at criticality, and exhibits a growing logarithmic discrepancy below it. A moving-saddle approximation also yields an asymptotically exact MFPT ratio under general strictly positive, finite-mean waiting times. Numerical comparisons across Gamma, Weibull, and generalized-Gamma renewal laws illustrate the approximation regimes and their finite-threshold accuracy. Together, these results provide an exact finite-threshold benchmark for the mean first-passage time and reveal how short-gap renewal statistics determine the asymptotic accuracy of high-threshold approximations.
\end{abstract}

\maketitle


\section{Introduction}
\label{sec:introduction}

In systems driven by random impulses, residual responses to earlier
inputs can overlap with subsequent pulses, making threshold-crossing
events depend on both pulse amplitudes and arrival timing
\cite{RichardsonSwarbrick2010}.
Exponentially relaxing shot noise captures this interplay, with each
positive impulse instantaneously increasing the signal, while its
contribution decays between successive arrivals
\cite{GilbertPollak1960,LaioEtAl2001}.
Consequently, a threshold may be crossed either by a single
sufficiently large impulse or through the accumulation of closely
spaced inputs \cite{DrosteLindner2017}.
In the latter case, crossing statistics can depend not only on the
mean arrival rate but also on the distribution of interarrival times
relative to the relaxation timescale
\cite{LyTranchina2009,DegerEtAl2012,Bremont2026}.
Such threshold dynamics arise in leaky integrate-and-fire neurons
\cite{HohnBurkitt2001,RichardsonSwarbrick2010} and in stochastic
hydrological models \cite{LaioEtAl2001,TameaEtAl2011}.
Bursty protein production provides another setting for impulsive
accumulation \cite{FriedmanCaiXie2006,KumarSinghKulkarni2015},
and the timing of some intracellular events triggered by protein
accumulation to a critical level can be described in terms of
first-passage times
\cite{GhusingaDennehySingh2017,CaoEtAl2019,RijalEtAl2022,HamEtAl2024}.
Across these settings, a central quantity of interest is the first-passage time (FPT), defined as the time at which a
signal first crosses a prescribed level
\cite{Masoliver1987,Redner2001,VanKampen2007,BrayMajumdarSchehr2013,
ChouDOrsogna2014,Redner2023}.
Its distribution and moments characterize the timing and variability
of these threshold events
\cite{HanggiTalknerBorkovec1990,Burkitt2006,BressloffNewby2013,
IyerBiswasZilman2016,Schwalger2021,QiuEtAl2020}.
These observations raise the central question of how the statistical
properties of pulse arrivals affect the mean first-passage time.

Classical Poisson shot noise provides a natural starting point for
first-passage analysis.
In this setting, successive interarrival times are independent and
exponentially distributed.
With independent pulse amplitudes and exponential relaxation, this
memoryless arrival law makes the shot-noise signal Markovian.
This Markovian structure facilitates the use of backward and
integral equations to analyze first-passage distributions
and their moments
\cite{TsuruiOsaki1976,Tuckwell1976,HernandezGarciaEtAl1987,
BorovkovNovikov2001}.
For exponentially distributed pulse amplitudes, explicit
first-passage transforms and MFPT expressions have been obtained
for filtered Poisson processes
\cite{PerryStadjeZacks2001,LaioEtAl2001,NovikovEtAl2005}.
Related threshold-crossing problems have been examined in
storage and shot-noise queueing models
\cite{KellaStadje2001,BoxmaMandjes2021}, while martingale
methods yield exit-time identities and high-threshold
asymptotics for L\'evy-driven Ornstein--Uhlenbeck processes
with exponentially distributed positive jumps
\cite{BorovkovNovikov2008}.
In many applications, however, interarrival statistics depart
from the exponential law.
In neuronal firing, refractory effects suppress short interspike intervals \cite{DegerEtAl2012}, while promoter regulation can produce nonexponential intervals between
transcriptional bursts \cite{HarperEtAl2011,KumarSinghKulkarni2015}.
Similar departures from Poisson timing occur in clustered
rainfall events \cite{Cowpertwait2001} and bursty packet
arrival streams \cite{PaxsonFloyd1995}.
For an exponentially relaxing signal, these differences in
arrival timing can alter the residual contributions from
earlier pulses, thereby affecting their accumulation and
the mean first-passage time.

\begin{figure*}[t]
\centering
\includegraphics[width=1\textwidth]{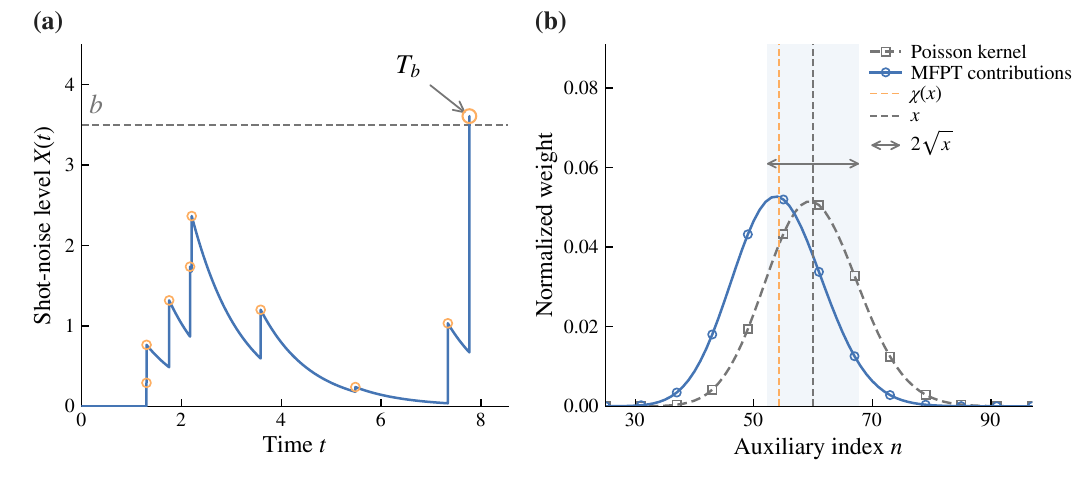}
\caption{Renewal shot-noise first passage and auxiliary Poisson averaging. (a) A sample trajectory starting from $X(0)=0$ with a fresh renewal interval. Blue curves show exponential relaxation between arrivals, and vertical segments indicate instantaneous jumps from $X(t_n^-)$ to $X(t_n)$. Open orange circles mark the post-arrival levels $X(t_n)=R_n/\lambda$. The larger circle marks the first crossing of $b=3.5$ at time $T_b$.
(b) Auxiliary Poisson probabilities $\mathbb{P}(N_x=n)$ (gray dashed line and squares) and normalized contributions $\widetilde C_n(x)$ to the exact MFPT series (blue solid line and circles) for $x=60$. The shaded interval $[x-\sqrt{x},x+\sqrt{x}]$ covers one standard deviation on either side of the Poisson mean, and the double-headed arrow marks its width $2\sqrt{x}$. Vertical dotted lines mark the Poisson mean $x$ and the continuous saddle $\chi(x)$. Renewal weighting shifts the dominant MFPT contributions below the Poisson mean. Both panels use unit-mean Gamma waiting times with shape $\kappa=0.5$, exponential pulse amplitudes, and $\lambda=\gamma=r=1$.}
\label{fig:model}
\end{figure*}

To further study these timing effects, we consider renewal shot noise, which extends the Poisson model by allowing successive waiting times to follow a general independent and identically distributed law rather than restricting them to an exponential distribution \cite{DassiosJangZhao2015,IksanovKabluchkoMarynych2016}.
Previous studies of related renewal shot-noise models
have established distributional limits
\cite{IksanovMarynychMeiners2014}, quantified variability
\cite{RajdlLansky2015}, derived conditional and asymptotic
moments \cite{JangDassiosZhao2018}, and examined tail
behavior under subexponential pulse amplitudes
\cite{Chen2019}. For time-dependent Poisson input, filtered models have also been
used to characterize joint cumulants and subthreshold
membrane-potential distributions \cite{BrighamDestexhe2015}.
These studies characterize statistical properties of the signal, whereas first-passage statistics depend on its evolution up to threshold crossing. More recently, Br\'emont derived a high-threshold product formula for exponentially marked renewal shot noise \cite{Bremont2026}. 
Equation~(1) of that work uses ratio asymptotics, while its short-gap expansion treats \(\kappa\le1/2\) separately. It also establishes an exponential high-threshold FPT limit. This motivates us to examine the relative accuracy of the fixed-index product against an exact finite-threshold MFPT benchmark. More precisely, we ask under what conditions evaluating the product at a deterministic index reproduces the MFPT with vanishing relative error and what corrections arise otherwise.

In this work, we study renewal shot noise with exponential relaxation at rate \(\gamma\) and exponentially distributed pulse amplitudes with rate parameter \(\lambda\). The signal starts from \(X(0)=0\) at the beginning of a fresh renewal interval. By solving a backward equation for the chain observed immediately after arrivals, we derive an exact series for the MFPT at any finite threshold and express it as an average over an auxiliary Poisson index with mean \(x=\lambda b\). This representation reveals a competition between the renewal-induced displacement of the dominant series terms and the fluctuation width of the Poisson index. For waiting-time densities with \(w(t)\sim ct^{\kappa-1}\)
as \(t\downarrow0\), the continuous saddle \(\chi(x)\) lies to the left of the Poisson mean \(x\), with
\(x-\chi(x)\sim a x^{1-\kappa}\) and
\(a=c\Gamma(\kappa)/\gamma^\kappa\).
We compare this shift with the Poisson standard deviation \(\sqrt{x}\).
For \(\kappa>1/2\), the shift becomes negligible relative to \(\sqrt{x}\) as \(x\to\infty\), and the ratio of the
exact MFPT to the fixed-index MFPT approximation tends
to one.
At \(\kappa=1/2\), the two scales are comparable, and the
ratio tends to \(e^{a^2/2}\).
However, for \(0<\kappa<1/2\), the shift exceeds the Poisson
fluctuation scale and the logarithm of the ratio is
asymptotic to \((a^2/2)x^{1-2\kappa}\).
Beyond this short-gap class, we also establish a moving-saddle MFPT equivalent and an exponential limit for the first-passage time normalized by its exact mean under general strictly positive, finite-mean waiting times, without
requiring a density or a short-gap power law for interarrival times. Furthermore, we obtain an absolutely convergent series for the
multiplicative correction relating the finite-threshold MFPT to the rare-event estimate. An explicit geometric bound on the remainder specifies how many terms suffice to achieve a prescribed absolute truncation error.
Event-driven simulations for Gamma and Weibull waiting
times provide an independent check of the exact MFPT
series. Using this series as a benchmark, we perform numerical
comparisons for Gamma, Weibull, and generalized-Gamma
renewal laws to illustrate the accuracy crossover and
quantify finite-threshold errors in the moving-saddle
approximation.

Specifically, our paper is organized as follows. Sec.~\ref{sec:route} derives the exact finite-threshold MFPT and expresses it as an auxiliary Poisson-index average. Sec.~\ref{sec:crossover_scale} identifies the crossover scale by comparing the renewal-induced displacement of the dominant series terms with the Poisson fluctuation width. Sec.~\ref{sec:regimes} develops the large-threshold theory. Specifically, Sec.~\ref{sec:stationary_bridge} connects stationary upcrossings to the MFPT from an initially empty state and derives a moving-saddle asymptotic equivalent; Sec.~\ref{sec:accuracy_regimes} establishes three regimes of accuracy for the deterministic-index product; and Sec.~\ref{sec:exp_limit} provides convergence of the normalized first-passage time to an exponential distribution without requiring a short-time power-law assumption. Sec.~\ref{sec:finite} compares exact MFPTs with direct event-driven simulations for Gamma and Weibull waiting times and examines the finite-threshold accuracy of the moving-saddle approximation across three waiting-time families. Finally, we discuss the implications and scope of these results in Sec.~\ref{sec:discussion}. Mathematical proofs and numerical procedures are given in Appendices~\ref{app:exact}--\ref{app:numerics}.

\section{Exact finite-threshold formula and auxiliary Poisson averaging}
\label{sec:route}
We consider a renewal shot-noise signal with
exponential relaxation,
\begin{equation}
X(t)=\sum_{t_j\le t}A_j e^{-\gamma(t-t_j)},
\qquad \gamma>0,
\end{equation}
where the arrival epochs are
\(t_j=\sum_{k=1}^{j}W_k\), \(j\ge1\).
Throughout this paper, the interarrival times
\(\{W_j\}_{j\ge1}\) are i.i.d.\ copies of a
strictly positive random variable \(W\) with
finite mean \(m=\E W\in(0,\infty)\).
Their common distribution is otherwise arbitrary,
and we write \(r=m^{-1}\) for the mean arrival rate. The marks \(\{A_j\}_{j\ge1}\) are i.i.d.\
exponential random variables, independent
of the waiting times, with
\(\Pp(A_j>u)=e^{-\lambda u}\) for \(u\ge0\)
and \(\lambda>0\).

We take \(X(0)=0\) and start at the beginning
of a fresh renewal interval, so that the
first impulse occurs at \(t_1=W_1>0\).
For a prescribed threshold \(b>0\), we define
the first-passage time
\begin{equation}
T_b=\inf\{t\ge0:X(t)>b\},
\end{equation}
and we introduce the dimensionless threshold
\(x=\lambda b\). Throughout the main text, \(T_b\) refers to
the empty initial state \(X(0)=0\) and the
notation \(T_b^{(z)}\) will be used below
for a nonzero initial level \(z\).
Fig.~\ref{fig:model}(a) shows a representative
trajectory of \(X(t)\), with upward jumps at
renewal epochs, exponential decay between
arrivals, and the first upcrossing of the
threshold \(b\).

\subsection{Exact finite-threshold means}
\label{sec:exact}

To obtain the exact MFPT at a finite threshold,
we examine the signal immediately after each
arrival. Since \(X(t)\) decreases between arrivals,
a threshold can only be crossed when a new
mark is added. We define the rescaled
post-arrival chain \(R_n=\lambda X(t_n^+)\),
with \(R_0=0\).

Setting \(Q_n=e^{-\gamma W_n}\) and
\(E_n=\lambda A_n\sim\Exp(1)\), we obtain the affine recursion
\begin{equation}
R_n=Q_nR_{n-1}+E_n,
\end{equation}
where the pairs \((Q_n,E_n)\) are independent and identically
distributed, with independent components.
Let
\(\nu_x=\inf\{n\ge1:R_n>x\}\)
denote the first-crossing index of the
embedded chain \(\{R_n\}_{n\ge0}\). Since crossing occurs at an arrival,
\begin{equation}
T_b=\sum_{j=1}^{\nu_x}W_j
\end{equation}
Moreover, the event that the \(j\)th waiting interval is reached
is determined by the preceding arrivals. The renewal stopping
identity therefore yields
\(\E T_b=m\E\nu_x=r^{-1}\E\nu_x\).

To evaluate \(\E\nu_x\), we introduce the Laplace transform
\(q(s)=\E e^{-\gamma sW}\) of the rescaled waiting time \(\gamma W\)
and set \(\phi(s)=1-q(s)\).
We also define \(\Pi_0=1\) and
\(\Pi_n=\prod_{j=1}^n[1-q(j)]\).
For \(0\le y<x\), let \(u_x(y)=\E_y\nu_x\) be the mean hitting
index when the embedded chain starts from \(R_0=y\) at the
beginning of a fresh renewal interval.
Since \(R_n\ge E_n\), an arrival with \(E_n>x\) necessarily
causes a crossing, which gives the geometric bound
\(u_x(y)\le e^x\).
Conditioning on the first waiting interval and the associated
mark, we obtain the backward equation
\begin{equation}
u_x(y)=1+\E_Q\left[\int_{Qy}^x e^{-(z-Qy)}u_x(z)\dd z\right].
\label{eq:backward}
\end{equation}
Here \(\E_Q\) denotes expectation with respect to the law of
\(Q=e^{-\gamma W}\).

To solve this equation, we introduce
\(H(y)=\int_y^x e^{-(z-y)}u_x(z)\,\dd z\)
and \(V(y)=H(0)-H(y)\).
With this change of variables, Eq.~\eqref{eq:backward}
reduces to the nonlocal initial-value problem
\begin{equation}
V'(y)=1+V(y)-\E_Q[V(Qy)],
\qquad V(0)=0.
\label{eq:nonlocal}
\end{equation}
Its unique solution is
\(V(y)=\sum_{n=1}^{\infty}(y^n/n!)\Pi_{n-1}\),
where the coefficients obey
\(\Pi_n=\Pi_{n-1}[1-q(n)]\).
Since \(H(x)=0\), it follows that \(u_x(0)=1+V(x)\).
Combining this identity with
\(\E T_b=r^{-1}\E\nu_x\), we obtain the exact MFPT
for every finite threshold
\begin{equation}
\E T_b
=
\frac1r\left[
1+\sum_{n=1}^{\infty}\frac{x^n}{n!}\Pi_{n-1}
\right].
\label{eq:exact_MFPT}
\end{equation}
Since $0<\Pi_n\le1$, the series converges
absolutely for every finite $x\ge0$
by comparison with the exponential series. Retaining its first two terms for a lower bound and comparing
the full series with \(e^x\) for an upper bound gives
\((1+x)/r\le\E T_b\le e^x/r\). Without relaxation, all previous inputs are retained,
and the exponential amplitude law gives a mean of
\(1+x\) arrivals before crossing. The upper bound instead follows by waiting for the first
impulse with \(A>b\), whose probability at each arrival is \(e^{-x}\). Residual signal can enable a smaller impulse to complete the crossing, reducing the mean below this single-impulse waiting time. The difference \(e^x/r-\E T_b\) thus measures how much temporal accumulation advances the mean crossing time.
An analogous single-impulse mechanism underlies the sparse-input firing rate of Poisson-driven leaky integrate-and-fire neurons \cite{DrosteLindner2017}.

At a fixed mean arrival rate, the waiting-time law
determines how much signal survives between inputs.
This information enters through
\(q(j)=\E[e^{-j\gamma W}]\), the moments of the retained
fraction \(Q=e^{-\gamma W}\), and hence through the
products \(\Pi_n\).
The exact series can therefore compare mean crossing
times for different arrival laws at a prescribed finite
threshold.
Such timing effects also appear in renewal-driven
integrate-and-fire models, where changing input regularity
at a fixed mean rate changes the first-crossing-time
distribution \cite{LyTranchina2009}.

The preceding derivation also extends to a nonzero
initial signal, provided the process starts at the
beginning of a fresh renewal interval.
For an initial level \(0\le z<b\), the process takes
the form
\begin{equation}
X_z(t)=ze^{-\gamma t}
+\sum_{j\ge1:t_j\le t}A_je^{-\gamma(t-t_j)}.
\label{eq:initial-process}
\end{equation}
\begin{figure*}[t]
\centering
\includegraphics[width=0.98\textwidth]{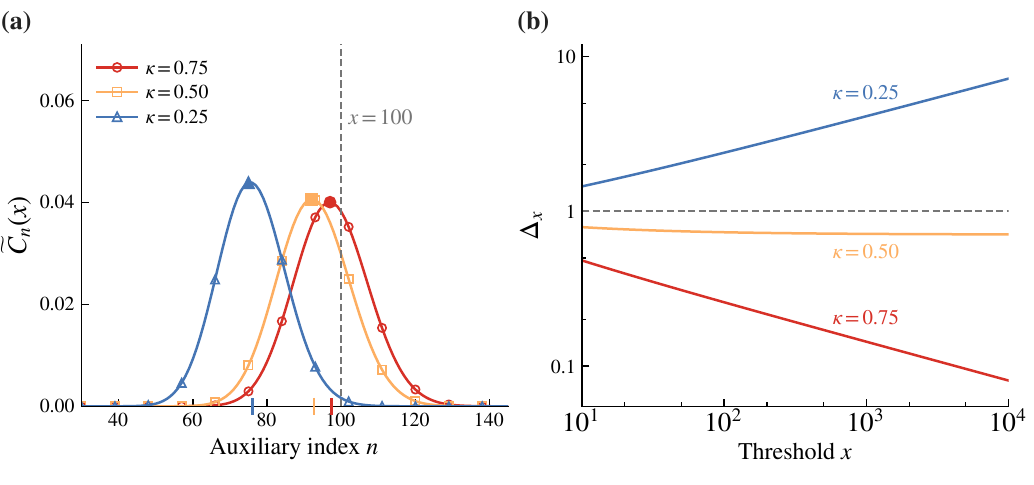}
\caption{Exact MFPT-series contributions and saddle displacement relative to the Poisson width.
(a) Normalized contributions $\widetilde C_n(x)$ to the exact MFPT series at $x=100$ for unit-mean Gamma waiting times with shapes $\kappa=0.75$, $0.50$, and $0.25$ (circles, squares, and triangles, respectively). Filled symbols mark the discrete maximizers $n_*(x)$, while short ticks along the horizontal axis indicate the continuous saddles $\chi(x)$. The vertical dotted line marks $x=100$.
(b) Saddle displacement $\Delta_x$ in units of the Poisson standard deviation [Eq.~\eqref{eq:Delta}]. The horizontal dotted line at $\Delta_x=1$ marks a displacement of one Poisson standard deviation. Both panels use exponential marks and $\lambda=r=\gamma=1$.}
\label{fig:crossover}
\end{figure*}
Let \(T_b^{(z)}\) denote the first-passage
time for an initial level \(X_z(0)=z\).
The notation is consistent with the
empty initial state used above, for which
\(T_b=T_b^{(0)}\).
Setting \(y=\lambda z\) and \(x=\lambda b\), we obtain
\begin{equation}
\E T_b^{(z)}
=\frac1r\left[
1+\sum_{n\ge1}\frac{\Pi_{n-1}}{n!}
\{x^n-q(n)y^n\}
\right].
\label{eq:initial-mean}
\end{equation}
Subtracting this expression from the empty-state
MFPT \( \E T_b^{(0)}=\E T_b \) yields
\begin{equation}
\E T_b^{(0)}-\E T_b^{(z)}
=\frac1r\sum_{n\ge1}
\frac{q(n)\Pi_{n-1}(\lambda z)^n}{n!}.
\label{eq:initial-offset}
\end{equation}
For fixed \(z\ge0\), the right-hand side is independent of
\(b>z\). As \(\E T_b^{(0)}\ge(1+x)/r\to\infty\), this implies
\(\E T_b^{(z)}/\E T_b^{(0)}\to1\) as \(b\to\infty\).

\subsection{The auxiliary Poisson transform}
\label{cor:poisson}
The result for nonzero initial conditions shows that a fixed initial level does not affect the leading
high-threshold behavior of the MFPT.
We therefore return to the empty initial state
and recast the exact series in
Eq.~\eqref{eq:exact_MFPT} as an expectation
over an auxiliary Poisson random variable. 

Let \(N_x\sim\operatorname{Pois}(x)\) be an auxiliary
random variable, and set \(B_0=1\) and
\(B_n=\Pi_{n-1}\) for \(n\ge1\).
Using the Poisson probability mass function,
Eq.~\eqref{eq:exact_MFPT} takes the exact form
\begin{equation}
\E T_b=\frac{e^x}{r}\E B_{N_x}.
\label{eq:Poisson_smoothing}
\end{equation}
This representation separates the Poisson averaging weights from the renewal-dependent factors $B_n$, providing a basis for analyzing the contributions of individual terms to the exact MFPT series. To identify which terms contribute most to the exact series,
define \(C_0(x)=1\) and
\(C_n(x)=x^n\Pi_{n-1}/n!\) for \(n\ge1\).
Normalizing these terms and using the Poisson representation, we obtain
\begin{equation}
\widetilde C_n(x)
:=\frac{C_n(x)}{\sum_{j\ge0}C_j(x)}
=\frac{\Pp(N_x=n)B_n}{\E[B_{N_x}]},
\end{equation}
where \(\widetilde C_n(x)\) is the fraction of the exact
MFPT series contributed by its \(n\)th term.
If the factors \(B_n\) were all equal to one, these weights
would reduce to the Poisson probabilities, whose largest
values occur near their mean \(x\).
In the renewal model, however, \(B_n\) is nonincreasing,
so the additional weighting favors smaller indices and may
shift the largest terms away from \(x\).
In Fig.~\ref{fig:model}(b), we compare the original Poisson weights $\Pp(N_x=n)$ with the normalized contributions $\widetilde C_n(x)$ to the exact MFPT series. This comparison shows how the renewal-dependent factors $B_n$ favor smaller indices and shift the largest series contributions away from the Poisson center $x$.

Given that \(N_x\) has mean and variance \(x\),
the Poisson probabilities are concentrated around
\(x\) on a scale of order \(\sqrt{x}\) ~\cite{Glynn1987,Ouimet2021}.
We use this scale to measure how far the indices
of the largest terms in the exact series are
shifted from \(x\).
In the next section, we quantify this displacement
and assess its relevance to the accuracy of
approximating the Poisson average by the
renewal-dependent product evaluated at a single index.


\section{Short-gap renewal statistics set the crossover scale}
\label{sec:crossover_scale}
The normalized contributions \(\widetilde C_n(x)\)
introduced in Sec.~\ref{cor:poisson} identify
the relative weight of each term in the exact
MFPT series. Since the normalization factor
is independent of \(n\), \(\widetilde C_n(x)\)
and \(C_n(x)\) have the same maximizers.
We therefore examine the ratio of successive
terms to locate the largest contributions.
For \(n\ge1\),
\begin{equation}
\frac{C_{n+1}(x)}{C_n(x)}
=
\frac{x}{n+1}[1-q(n)].
\label{eq:summand_ratio}
\end{equation}
The first factor in Eq.~\eqref{eq:summand_ratio}
gives the ratio of successive Poisson weights,
whereas the second encodes the dependence on
the renewal waiting-time distribution.
Since \(C_n(x)>0\), we have
\begin{equation}\label{eq:successive terms}
\begin{aligned}
\frac{C_{n+1}(x)}{C_n(x)}>1
&\iff C_{n+1}(x)>C_n(x),\\
\frac{C_{n+1}(x)}{C_n(x)}<1
&\iff C_{n+1}(x)<C_n(x).
\end{aligned}
\end{equation}
By Eq.~\eqref{eq:successive terms}, the largest
terms in the sequence \(\{C_n(x)\}_{n\ge1}\)
occur near the index at which the ratio of
successive terms drops below unity.
Formally setting the ratio in
Eq.~\eqref{eq:summand_ratio} equal to unity
yields the balance condition
\(n+1=x[1-q(n)]\).
For large indices, replacing \(n+1\) by \(n\)
and treating the index as a continuous variable
leads to the approximate balance equation
\(s=x[1-q(s)]\).

For \(x>r/\gamma\), we denote the positive solution
of this equation by \(\chi(x)\), so that
\begin{equation}
\frac{\chi(x)}{1-q(\chi(x))}=x.
\label{eq:balance_eq}
\end{equation}
In the large-threshold regime, \(\chi(x)\) provides
a continuous approximation to the index of the
largest summand in the exact MFPT series
\(\E T_b=r^{-1}\sum_{n=0}^{\infty}C_n(x)\).
The difference \(x-\chi(x)\) then measures the
displacement of this approximate location from
the Poisson center \(x\).
At finite \(x\), however, \(\chi(x)\) need not
coincide with the exact discrete maximizer
\(n_*(x):=\min\operatorname*{arg\,max}_{n\ge0}C_n(x)\).
If the maximum is attained at more than one index, we select
the smallest maximizing index.
Rearranging Eq.~\eqref{eq:balance_eq}, we obtain
\begin{equation}
x-\chi(x)=xq(\chi(x)).
\end{equation}
Since \(q(s)>0\) for \(s>0\), the continuous
saddle lies to the left of the Poisson center \(x\).
As \(x\to\infty\), the positive solution
\(\chi(x)\to\infty\). Since \(q(s)\to0\) as
\(s\to\infty\), the balance equation implies
\(\chi(x)/x=1-q(\chi(x))\to1\), and hence
\(\chi(x)\sim x\).

To determine the large-threshold behavior of
the saddle displacement, we consider waiting-time
densities satisfying \(w(t)\sim c t^{\kappa-1}\)
as \(t\downarrow0\), with \(c>0\) and \(\kappa>0\).
This short-time condition implies
\(q(s)\sim a s^{-\kappa}\) as \(s\to\infty\),
where \(a=c\Gamma(\kappa)/\gamma^\kappa\).
Combining this asymptotic relation with
\(\chi(x)\sim x\), we obtain
\(x-\chi(x)\sim a x^{1-\kappa}\).

Since \(N_x\sim\operatorname{Pois}(x)\), the auxiliary
Poisson index fluctuates around \(x\) on a scale
of order \(\sqrt{x}\).
The exact Poisson representation in
Eq.~\eqref{eq:Poisson_smoothing} therefore averages
the renewal-dependent factors \(B_n\) with
Poisson weights concentrated on this scale.
To compare the saddle displacement \(x-\chi(x)\)
with the Poisson fluctuation scale \(\sqrt{x}\),
we introduce the dimensionless ratio
\begin{equation}
\Delta_x
:=\frac{x-\chi(x)}{\sqrt{x}}
=\sqrt{x}q(\chi(x))
\sim a x^{1/2-\kappa}.
\label{eq:Delta}
\end{equation}

To relate this displacement scale to the accuracy
of the single-index approximation of the Poisson
average in Eq.~\eqref{eq:Poisson_smoothing},
we examine how the renewal-dependent product
\(B_n\) varies across the Poisson fluctuation
window. For fixed \(u\ne0\), let
\(n=\lfloor x+u\sqrt{x}\rfloor\).
Since \(q(j)\sim q(x)\) uniformly for
\(j=x+O(\sqrt{x})\) and
\(\log[1-q(j)]\sim-q(j)\), the product
representation \(B_n=\Pi_{n-1}\) gives
\begin{equation}
\log\frac{B_n}{B_{\lfloor x\rfloor}}
\sim -u\sqrt{x}q(x)
\sim -u\Delta_x.
\label{eq:product_window_variation}
\end{equation}
The second asymptotic equivalence follows from
\(q(x)\sim q(\chi(x))\), which holds because
\(\chi(x)\sim x\).
Thus, \(\Delta_x\) also determines the asymptotic
scale of the logarithmic variation of the
renewal-dependent product across the Poisson
fluctuation window.

In Fig.~\ref{fig:crossover}, we compare the exact
summand profiles and examine how the normalized
displacement varies with the threshold.
As \(x\) increases, \(\Delta_x\) vanishes for
\(\kappa>1/2\), remains of order unity at
\(\kappa=1/2\), and diverges for
\(0<\kappa<1/2\).
In particular, for \(1/2<\kappa<1\), the
waiting-time density diverges at zero, yet the
saddle displacement becomes negligible relative
to \(\sqrt{x}\).
Thus, the boundary \(\kappa=1/2\) marks a change
in the displacement relative to the Poisson
fluctuation scale. This scale competition
underlies the crossover in the accuracy of a
sharp-index approximation to the MFPT, rather
than the onset of short-gap clustering.
The displacement illustrated in Fig.~\ref{fig:model}(b)
is further examined across waiting-time distributions
in Fig.~S2 of the Supplemental
Material~\cite{SupplementalMaterial}.
At the common threshold $x=60$, the auxiliary
Poisson kernel remains unchanged, whereas the
location and shape of the dominant MFPT-series
contributions vary with the renewal law and the
short-time exponent $\kappa$.

To connect this auxiliary-index analysis with
the physical MFPT, we next relate first passage
from an empty signal to stationary upcrossings.

\section{Asymptotic accuracy of the mean first-passage time}
\label{sec:regimes}

\subsection{From stationary upcrossings to the moving saddle}
\label{sec:stationary_bridge}

To connect the auxiliary-index displacement with the
physical MFPT, we relate first passage from an empty
signal to upcrossings of the stationary embedded chain.
Let \(R\) denote a stationary post-arrival level of
the embedded chain. Its stationary law is unique
and admits a continuous, strictly positive density
\(f_R\) on \((0,\infty)\). The stationary level \(R\) admits the perpetuity
representation \cite{DiaconisFreedman1999,GoldieMaller2000,
AlsmeyerIksanovRosler2009}
\begin{equation}
R\overset d=
\sum_{k=0}^{\infty}
\widehat E_{k+1}\prod_{j=1}^k\widehat Q_j,
\label{eq:perpetuity}
\end{equation}
where \(\{(\widehat Q_j,\widehat E_j)\}_{j\ge1}\)
are i.i.d.\ copies of \((Q,E)\).
Separating the first term of this series yields
the distributional fixed-point equation
\(R\overset d=QR'+E\),
where \(R'\overset d=R\) is independent of
the pair \((Q,E)\). Setting \(Z=QR'\) and using \(E\sim\Exp(1)\),
we obtain the stationary density
\begin{equation}
\begin{aligned}
f_R(x)
&=
\E[e^{-(x-Z)}\ind_{\{Z\le x\}}]\\
&=
\Pp\{Z\le x<Z+E\},
\end{aligned}
\label{eq:stationary_density}
\end{equation}
where the second equality follows by
conditioning on \(Z\).
Thus, \(f_R(x)\) equals the probability that
a single incoming exponential mark pushes
the stationary pre-arrival level \(Z\)
across the threshold \(x\).

For an initial post-arrival state \(U\), independent
of future innovations, let \(R_n^{(U)}\) denote
the corresponding embedded chain and define
\begin{equation}
u_n^{(U)}(x)
=\Pp\{Q_nR_{n-1}^{(U)}\le x<R_n^{(U)}\}.
\label{eq:upcrossing}
\end{equation}
When \(U\) is random, the probability includes averaging over its
initial distribution.
We write \(u_n^{(0)}(x)\) for \(U=0\) and
\(u_n^{(+)}(x)\) for \(U=Y_x=x+\widetilde E\),
where \(\widetilde E\sim\Exp(1)\) is independent
of the future innovations.

To relate the stationary upcrossing probability
to the mean first-crossing index, we introduce
\begin{equation}
K_x=1+\sum_{n\ge1}
[u_n^{(+)}(x)-u_n^{(0)}(x)].
\label{eq:K-def}
\end{equation}
Conditioning on the first upcrossing and using
the independent exponential overshoot, a
comparison of the resulting generating
functions gives the exact identity
\begin{equation}
f_R(x)\E_0\nu_x=K_x,
\label{eq:K-exact}
\end{equation}
where \(K_x\to1\) as
\(x\to\infty\). To establish this limit, we split
the upcrossing series at a logarithmic cutoff.
Strict positivity of the waiting times makes the
early-upcrossing contribution vanish, while coupling
the chains through common innovations gives a
geometrically decaying bound for the late-time
difference. The complete proof, including the
exact identity above, is given in
Appendix~\ref{app:Kx}.

Using Eq.~\eqref{eq:K-exact} and
\(\E T_b=r^{-1}\E_0\nu_x\), we obtain
\begin{equation}
\E T_b
=\frac{K_x}{r f_R(x)}
\sim\frac{1}{r f_R(x)},
\quad x=\lambda b\to\infty.
\label{eq:MFPT_Kx}
\end{equation}

Therefore, the large-threshold MFPT is
determined to leading order by the stationary
upcrossing density. This reduction concerns the mean and does not by itself establish a limit law for the first-passage-time distribution.
At a finite threshold, the correction to the rare-event estimate can be evaluated with an explicit truncation error bound. Writing $T_{\mathrm{RE}}(b)=[r f_R(x)]^{-1}$, the exact MFPT identity above gives
$\mathbb{E}T_b/T_{\mathrm{RE}}(b)-1=K_x-1$.
For an integer $L\ge 1$, define
\begin{equation}
K_{x,L}
=
1+\sum_{n=1}^{L}
\left[u_n^{(+)}(x)-u_n^{(0)}(x)\right].
\label{eq:K-partial-main}
\end{equation}
With $\rho=q(1)\in(0,1)$, the coupling argument in Appendix \ref{app:Kx}
yields
\begin{equation}
\left|K_x-K_{x,L}\right|
\le
\frac{(x+1)\rho^L}{1-\rho}
=:\tau_{x,L}.
\label{eq:K-truncation-main}
\end{equation}
The relative correction therefore lies in the explicit interval
\begin{equation}
K_{x,L}-1-\tau_{x,L}\le
\frac{\mathbb{E}T_b}{T_{\mathrm{RE}}(b)}-1
\le
K_{x,L}-1+\tau_{x,L}.
\label{eq:MFPT-correction-interval}
\end{equation}
Thus, evaluating the first $L$ terms of the correction series
determines the finite-threshold departure from the inverse
stationary upcrossing rate with an absolute truncation error
at most $\tau_{x,L}$. An explicit sufficient choice of $L$
for any prescribed tolerance is given in Appendix \ref{app:Kx}.
The bound holds for every finite threshold and every strictly
positive, finite-mean renewal waiting-time law considered here.

We next derive the large-\(x\) asymptotics
of the stationary density \(f_R(x)\)
under the general waiting-time assumptions
\(W>0\) almost surely and
\(m=\E W\in(0,\infty)\).
Recall that \(\phi(s)=1-q(s)\).
By dominated convergence,
\begin{equation}
q(s)\to0,\quad
s\phi'(s)
=\E[\gamma sWe^{-\gamma sW}]
\to0
\label{eq:general-phi}
\end{equation}
as $s\to\infty$.
Since \(\phi(s)-s\phi'(s)>0\) for \(s>0\),
the function \(g(s)=s/\phi(s)\) is strictly
increasing. Moreover, \(g(s)\to r/\gamma\)
as \(s\downarrow0\) and \(g(s)\to\infty\)
as \(s\to\infty\). Thus, \(g\) is a
bijection from \((0,\infty)\) onto
\((r/\gamma,\infty)\).
Its inverse coincides with the continuous
saddle coordinate \(\chi\) introduced in
Eq.~\eqref{eq:balance_eq}.
Eq.~\eqref{eq:general-phi} further
implies \(g(s)\sim s\) and \(g'(s)\to1\),
which yield \(\chi(x)\sim x\) and
\(\chi'(x)\to1\) as \(x\to\infty\).
The corresponding inverse function
estimates are established in the Appendix~\ref{app:saddle}.
\begin{figure*}[t]
\centering
\includegraphics[width=1\textwidth]{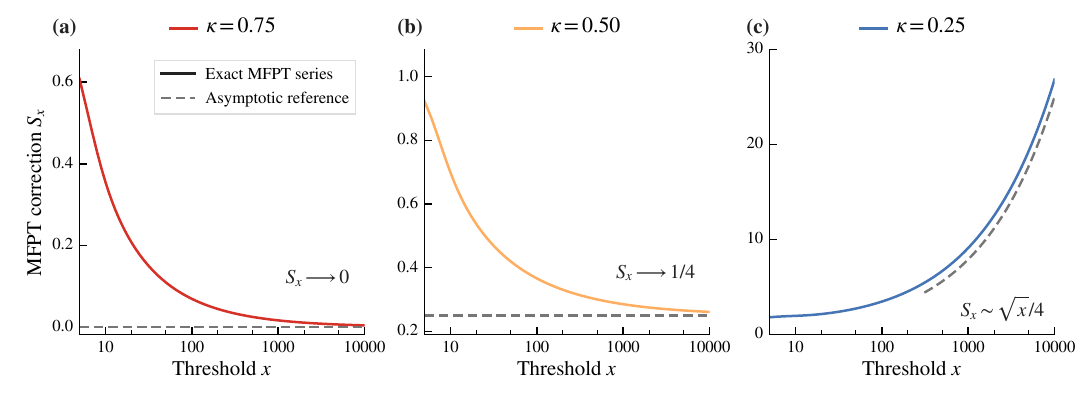}
\caption{Finite-threshold corrections to the fixed-index product approximation in three renewal regimes.
Solid curves show the logarithmic MFPT ratio $S_x=\log[\mathbb{E}T_b/T_{\mathrm{prod}}(b)]$, evaluated from the exact MFPT series at integer $x=\lambda b$, for unit-mean Gamma waiting times with shape $\kappa$. Dashed lines indicate the corresponding asymptotic limits or leading-order behavior. All panels use exponential marks and $\lambda=r=\gamma=1$.
(a) For $\kappa=0.75$, $S_x\to0$, and the fixed-index approximation is asymptotically exact in ratio.
(b) For $\kappa=0.50$, $S_x\to1/4$, corresponding to a limiting MFPT ratio of $e^{1/4}$.
(c) For $\kappa=0.25$, $S_x\sim\sqrt{x}/4$, giving the leading growth of the logarithmic MFPT ratio.}
\label{fig:regimes}
\end{figure*}
We next introduce an auxiliary compound-Poisson
subordinator ~\cite{Bertoin1999Subordinators,
BertoinVanHarnSteutel1999} $\{
\xi_t=\gamma\sum_{j=1}^{\mathcal N_t}\widehat W_j,
 t\ge0\},$
where \((\mathcal N_t)_{t\ge0}\) is a rate-one
Poisson process and
\(\{\widehat W_j\}_{j\ge1}\) are i.i.d.\
copies of \(W\), independent of
\(\mathcal N_t\). The Laplace exponent of the subordinator
\((\xi_t)_{t\ge0}\) is \(\phi(s)=1-q(s)\).
Its exponential functional is defined by
\begin{equation}
I=\int_0^\infty e^{-\xi_t}\,\dd t.
\end{equation}
Decomposing the integral over the successive
holding intervals of the auxiliary Poisson
process yields the same perpetuity
representation as Eq.~\eqref{eq:perpetuity}.
Consequently, \(R\overset d=I\).
Since \(R\overset d=I\) and
\(s\phi'(s)/\phi(s)\to0\), the auxiliary
subordinator satisfies condition (H) of
Haas~\cite{Haas2025} with Haas's inverse function identified as \(\chi\). Applying the corresponding
density asymptotics gives, for any fixed
\(x_0>r/\gamma\),
\begin{equation}
f_R(y)
=
C\sqrt{\chi'(y)}e^{-J(y)}[1+o(1)],
\quad y\to\infty,
\label{eq:Haas-density}
\end{equation}
where
\(J(y)=\int_{x_0}^{y}\frac{\chi(u)}{u}\dd u\)
and \(C>0\) is a constant independent of \(y\).
To identify the normalization constant in
terms of the renewal product, we use the
exact integer moments of the stationary level,
\begin{equation}
\E R^n=\frac{n!}{\Pi_n},
\qquad n\ge1.
\label{eq:stationary-moments}
\end{equation}
Their integral representation is
\begin{equation}
\E R^n=\int_0^\infty y^n f_R(y)\,\dd y.
\end{equation}
For the asymptotic evaluation of this
integral, Eq.~\eqref{eq:Haas-density}
identifies the Laplace-type exponent
\(F_n(y)=n\log y-J(y)\).
Since \(F_n'(y)=[n-\chi(y)]/y\),
the exponent has a unique maximum at
\(y_n=n/\phi(n)\), where \(\chi(y_n)=n\).
Applying Laplace's method to the moment
integral and matching the resulting
asymptotics with the exact moments in
Eq.~\eqref{eq:stationary-moments},
we obtain the asymptotic prefactor \begin{equation}
C=\lim_{n\to\infty}
\frac{n!\,e^{J(y_n)}}
{\Pi_n\sqrt{2\pi y_n}\,y_n^n}.
\end{equation}
Transferring the density asymptotics from
\(y_{n_x}\) to \(x\), with
\(n_x=\lfloor\chi(x)\rfloor\), we derive
\begin{equation}
f_R(x)\sim\Pi_{n_x}^{-1}
\exp\{n_x\log(n_x/x)-n_x\}.
\label{eq:density-moving-saddle}
\end{equation}
Taking the reciprocal of
Eq.~\eqref{eq:density-moving-saddle}
and combining it with Eq.~\eqref{eq:MFPT_Kx},
we obtain the moving-saddle asymptotic
equivalent for the MFPT.
With \(n_x=\lfloor\chi(x)\rfloor\)
and \(D(x,n)=x-n+n\log(n/x)\), this gives
\begin{equation}
\E T_b
\sim
\frac{e^x}{r}\Pi_{n_x}
\exp[-D(x,n_x)].
\label{eq:moving_saddle_MFPT}
\end{equation}
The product \(\Pi_{n_x}\) retains the
dependence on the renewal law, while
\(D(x,n_x)\) quantifies the penalty
associated with displacing the saddle
index from the Poisson center \(x\).

To quantify this penalty, we note that
\(D(x,x)=\partial_n D(x,x)=0\) and
\(\partial_n^2 D(x,n)=1/n\).
Taylor's theorem gives
\(D(x,\chi(x))=(x-\chi(x))^2/(2\xi_x)\),
where \(\xi_x\) lies between
\(\chi(x)\) and \(x\).
Since \(\chi(x)/x\to1\), we also have
\(\xi_x/x\to1\), and hence
\begin{equation}
D(x,\chi(x))
=\frac{\Delta_x^2}{2}[1+o(1)],
\qquad x\to\infty,
\label{eq:D_displacement}
\end{equation}
which relates the leading penalty to
the squared normalized saddle displacement.
The density localization, transfer estimates,
and control of the floor approximation are established
in Appendix~\ref{app:saddle}.
The finite-threshold relative accuracy of this equivalent
is examined in Sec.~\ref{sec:numerical_saddle}.

\subsection{Three regimes of asymptotic accuracy}
\label{sec:accuracy_regimes}

We now return to the short-gap power-law class introduced in
Sec.~\ref{sec:crossover_scale}.
In this class, the interarrival density
satisfies \(w(t)\sim c t^{\kappa-1}\)
as \(t\downarrow0\), with \(c>0\)
and \(\kappa>0\).

To this end, we compare the exact MFPT with the fixed-index product
\(T_{\mathrm{prod}}(b)=(e^x/r)\Pi_{\lfloor x\rfloor}\).
For integer \(x=\lambda b\), this is the right-hand side
of Eq.~(1) in Ref.~\cite{Bremont2026}; for noninteger
\(x\), the floor gives a specified integer-index
comparator. By the exact Poisson identity in
Eq.~\eqref{eq:Poisson_smoothing}, their ratio is
\begin{equation}
\frac{\E T_b}{T_{\mathrm{prod}}(b)}
=\frac{\E B_{N_x}}{\Pi_{\lfloor x\rfloor}}.
\label{eq:product_exact_ratio}
\end{equation}
Thus, the asymptotic comparison reduces to determining
whether the Poisson-index average agrees in ratio
with the product evaluated at a prescribed index.
We write
\(S_x:=\log[\E T_b/T_{\mathrm{prod}}(b)]\),
so that \(e^{S_x}\) is the corresponding MFPT ratio.

The moving-saddle equivalent
in Eq.~\eqref{eq:moving_saddle_MFPT}
gives the asymptotic decomposition
\begin{equation}
S_x
=
\log\frac{\Pi_{n_x}}{\Pi_{\lfloor x\rfloor}}
-D(x,n_x)+o(1),
\label{eq:log_accuracy_decomposition}
\end{equation}
where the first term accounts for the change in
the renewal product when its index moves
from \(\lfloor x\rfloor\) to \(n_x\) and
the second measures the corresponding
Poisson displacement penalty. For the short-gap class introduced in
Sec.~\ref{sec:crossover_scale},
\(q(s)\sim a s^{-\kappa}\).

For \(0<\kappa\le1/2\), the saddle equation
\(x-\chi(x)=xq(\chi(x))\), together with
\(n_x=\lfloor\chi(x)\rfloor\), gives
\(x-n_x\sim a x^{1-\kappa}\).
Since \(n_x/x\to1\), the relation
\(q(s)\sim a s^{-\kappa}\) implies
\(q(j)=q(x)[1+o(1)]\) uniformly for
\(n_x<j\le\lfloor x\rfloor\).
Expanding \(-\log[1-q(j)]\) and summing
over these indices, we find that the
logarithmic product difference is
asymptotic to
\((\lfloor x\rfloor-n_x)q(x)\).
Similarly, a Taylor expansion of
\(D(x,n)\) around \(n=x\) gives
\(D(x,n_x)\sim(x-n_x)^2/(2x)\).
Together, these estimates yield
\begin{equation}
\begin{aligned}
\log\frac{\Pi_{n_x}}{\Pi_{\lfloor x\rfloor}}
\sim a^2 x^{1-2\kappa}
\end{aligned}
\label{eq:correction_balance}
\end{equation}
and 
\begin{equation}\label{eq:correction_balance_D}
D(x,n_x)\sim\frac{a^2}{2}x^{1-2\kappa}.
\end{equation}
From Eq.~\eqref{eq:correction_balance} and Eq.~\eqref{eq:correction_balance_D}, we find
that the increase in the logarithm of the
renewal product is partially offset by
the Poisson displacement penalty,
leaving a leading logarithmic correction
of \(a^2x^{1-2\kappa}/2\). In particular, at the critical exponent
\(\kappa=1/2\), this correction approaches
the finite value \(a^2/2\).

For \(\kappa>1/2\), the saddle equation
and \(n_x=\lfloor\chi(x)\rfloor\) give
\(x-n_x=xq(\chi(x))+O(1)
=O(x^{1-\kappa}+1)\).
Since \(q(j)=O(x^{-\kappa})\) uniformly
for \(n_x<j\le\lfloor x\rfloor\),
the logarithmic product difference is
\(O(x^{1-2\kappa}+x^{-\kappa})=o(1)\).
A Taylor expansion of \(D(x,n)\) around
\(n=x\) similarly gives
\(D(x,n_x)=O(x^{1-2\kappa}+x^{-1})=o(1)\).
Thus, both correction terms in
Eq.~\eqref{eq:log_accuracy_decomposition}
vanish as \(x\to\infty\).

These estimates determine the relative accuracy of the fixed-index product in all three regimes
\begin{align}
\frac{\E T_b}{T_{\mathrm{prod}}(b)}&\longrightarrow1,
&&\kappa>\frac12,
\label{eq:regime_super}
\\
\frac{\E T_b}{T_{\mathrm{prod}}(b)}&\longrightarrow e^{a^2/2},
&&\kappa=\frac12,
\label{eq:regime_critical}
\\
\log\frac{\E T_b}{T_{\mathrm{prod}}(b)}&\sim\frac{a^2}{2}x^{1-2\kappa},
&&0<\kappa<\frac12.
\label{eq:regime_sub}
\end{align}
Accordingly, for \(\kappa>1/2\), the
fixed-index product remains asymptotically
exact in ratio, since both the renewal-product
gain and the Poisson displacement penalty
vanish in the large-threshold limit.
This contrasts with the critical case
\(\kappa=1/2\), where the two contributions
leave a finite multiplicative correction
\(e^{a^2/2}\).
Below the critical value, the renewal-product
gain still exceeds the displacement penalty,
but their difference grows on the logarithmic
scale, causing the fixed-index product to
underestimate the MFPT by an unbounded factor. Within this short-gap class, the fixed-index expression of Eq.~(1)
in Ref.~\cite{Bremont2026} is therefore asymptotically equivalent
to the MFPT only for \(\kappa>1/2\), whereas its ratio to the exact
mean does not approach unity at or below the crossover.
\begin{figure*}[t]
\centering
\includegraphics[width=0.97\textwidth]{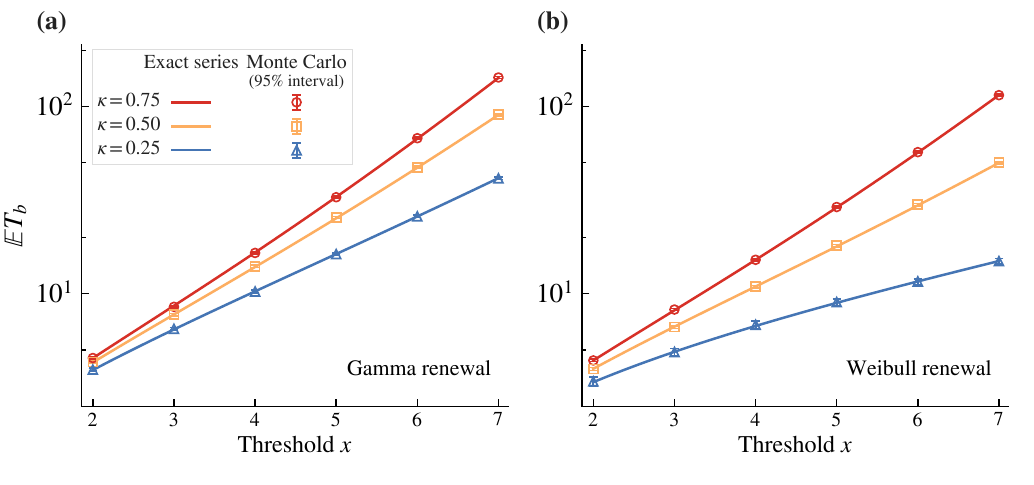}
\caption{Finite-threshold mean first-passage times for Gamma and Weibull renewal processes.
Waiting times have unit mean and follow (a) Gamma or (b) Weibull distributions. Solid curves show numerical evaluations of the exact MFPT series. Open circles, squares, and triangles denote event-driven Monte Carlo estimates for $\kappa=0.75$, $0.50$, and $0.25$, respectively. Error bars indicate pointwise approximate $95\%$ confidence intervals for the Monte Carlo means, based on the normal approximation. For each waiting-time family and $\kappa$, the same ensemble of 22,000 independent trajectories is used at all six thresholds $x=2,3,4,5,6,7$, so estimates across thresholds are correlated. All trajectories start from $X(0)=0$ with a fresh waiting interval. Pulse amplitudes are exponentially distributed, with $\lambda=r=\gamma=1$.}
\label{fig:exact}
\end{figure*}
The three regimes are illustrated in
Fig.~\ref{fig:regimes}, which shows
\(S_x\) evaluated from the exact MFPT
series for mean-one Gamma renewal
intervals. As \(x\) increases, the
logarithmic correction decreases toward
zero, approaches a finite plateau, or
continues to grow, consistent with the
asymptotic behavior derived above.
The normalized comparisons in Fig.~S1 of the
Supplemental Material~\cite{SupplementalMaterial}
show the same accuracy classification for
mean-one Gamma, Weibull, and generalized-Gamma
renewal intervals, while retaining their distinct
finite-threshold corrections.
A more refined critical asymptotic,
based on a higher-order short-time
expansion of the waiting-time density,
is derived in Appendix~\ref{app:refined}.

\subsection{Exponential limit of the normalized first-passage time}
\label{sec:exp_limit}

The accuracy crossover concerns the MFPT
timescale rather than the limiting form of
the normalized first-passage time distribution.
Br\'emont~\cite{Bremont2026} previously reported
an exponential high-threshold limit for
renewal shot noise with interarrival times
admitting a density.
Here, we establish the corresponding limit
for general i.i.d.\ strictly positive
waiting times with finite mean, without
requiring their distribution to admit a density.

For a signal initially empty at the beginning
of a fresh renewal interval, with independent
exponential marks, we obtain
\begin{equation}
\frac{T_b}{\E T_b}\Rightarrow\Exp(1),
\qquad b\to\infty.
\label{eq:exp_limit}
\end{equation}
The proof follows from the exact
finite-threshold Laplace transform derived
in Appendix~\ref{app:laplace}, evaluated
at the scale \(s=z/\E T_b\). Unlike the accuracy classification, this limit requires no short-time
power-law assumption.
Thus, although the renewal law affects
the mean rare-crossing timescale,
the first-passage time retains an
exponential limiting distribution after
normalization by its exact mean.
An exponential high-threshold limit was
previously established for
L\'evy-driven Ornstein--Uhlenbeck processes
with exponentially distributed positive jumps
\cite{BorovkovNovikov2008}.
More generally, first-passage limits for rare sets have
been studied in regenerative processes
\cite{GlassermanKou1995}, while exponential hitting-time
laws for positive recurrent Harris chains and processes
were established in Ref.~\cite{Glynn2011}.

\section{Finite-threshold MFPTs and moving-saddle accuracy}
\label{sec:finite}
\subsection{Gamma and Weibull renewal laws}
\label{sec:finite_benchmarks}
To examine first-passage behavior at finite thresholds,
we use the exact MFPT series to compare Gamma and
Weibull renewal laws, for which high-threshold
asymptotic approximations may not yet be accurate.
For both families, we choose mean-one waiting times
with \(\kappa=0.75\), \(0.50\), and \(0.25\), and set
\(\lambda=r=\gamma=1\).
Fig.~\ref{fig:exact} displays the corresponding
MFPTs calculated from Eq.~\eqref{eq:exact_MFPT}.
Within each family, the MFPT increases rapidly with
the threshold, while smaller values of \(\kappa\)
yield shorter crossing times over the range examined.
This trend is consistent with a greater prevalence
of closely spaced arrivals. However, varying
\(\kappa\) also changes the waiting-time distribution
away from the origin, so the observed differences
cannot be attributed to short-time behavior alone.

Comparing the two families at fixed \(\kappa\)
further reveals the influence of the full renewal
law. At \(\kappa=1/2\) and \(x=7\), for example,
the exact MFPT is approximately \(90.2\) for
Gamma waiting times, compared with \(50.0\)
for Weibull waiting times. Despite having the
same mean waiting time and short-gap exponent,
these distributions differ in both the coefficient
of the short-time power law and their behavior
at longer intervals. Such differences are
reflected in the Laplace transform \(q(s)\)
and consequently enter the exact MFPT series.
Hence, although \(\kappa\) determines the
asymptotic accuracy regime identified in
Sec.~\ref{sec:accuracy_regimes}, it does not
uniquely determine the MFPT at finite thresholds.

Moreover, to check the exact-series predictions independently,
we perform direct event-driven simulations of
the continuous-time shot-noise process. Open symbols
in Fig.~\ref{fig:exact} represent the simulated
mean first-passage times, which are consistent
with the exact-series curves for both renewal
families and all three values of \(\kappa\)
at \(x=2,\ldots,7\). By contrast, the
high-threshold correction in Fig.~\ref{fig:regimes}
is assessed using deterministic evaluations of
the series.
Details of the sampling procedure and the
dependence between estimates at different
thresholds are provided in
Appendix~\ref{app:numerics}.

\subsection{Accuracy of the moving-saddle approximation}
\label{sec:numerical_saddle}
\begin{figure*}[t]
\centering
\includegraphics[width=1\textwidth]{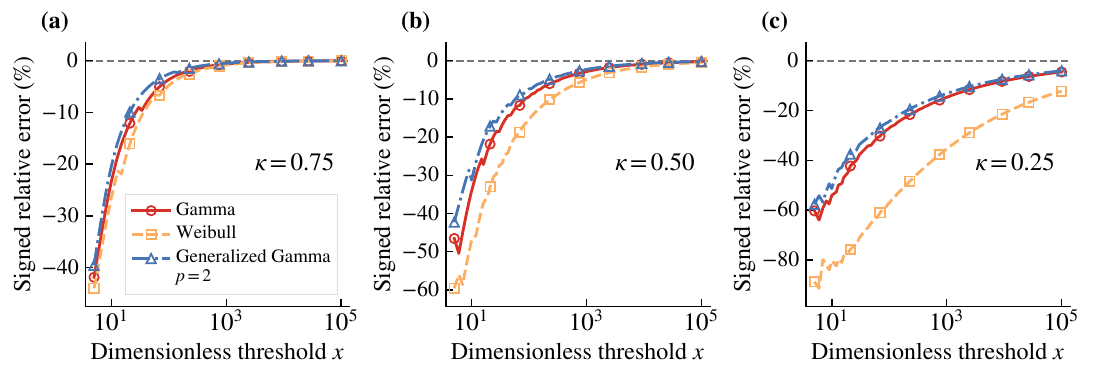}
\caption{Finite-threshold accuracy of the moving-saddle MFPT approximation.
Signed percentage errors $100\varepsilon_{\mathrm{sad}}(x)$ relative to the exact MFPT are shown for (a) $\kappa=0.75$, (b) $\kappa=0.50$, and (c) $\kappa=0.25$. Each panel compares unit-mean Gamma (red solid lines, open circles), Weibull (orange dashed lines, open squares), and generalized-Gamma waiting times with $p=2$ (blue dash-dotted lines, open triangles). Lines and symbols represent deterministic numerical evaluations. Gray dotted lines mark zero error, with negative values indicating underestimation of the exact MFPT. The errors approach zero as $x$ increases, with slower convergence for $\kappa=0.25$. All panels use exponential marks and $\lambda=\gamma=r=1$.}
\label{fig:moving_saddle_accuracy}
\end{figure*}
The moving-saddle equivalent in
Eq.~\eqref{eq:moving_saddle_MFPT} applies to general
strictly positive renewal waiting times with finite mean. For numerical evaluation, we restrict to thresholds satisfying
\(n_x=\lfloor\chi(x)\rfloor\ge1\), where the saddle expression
is well defined. 
To examine its finite-threshold accuracy, we compare
it with the exact mean $M_x=\E T_b$ for
mean-one Gamma, Weibull, and generalized-Gamma
waiting times with $p=2$.
For each waiting-time law, the saddle coordinate
$\chi(x)$ and the product $\Pi_{n_x}$ are determined
from its full Laplace transform $q(s)$.
We measure the discrepancy by the signed relative error
\begin{equation}
\varepsilon_{\mathrm{sad}}(x)
:=\frac{M_{\mathrm{sad}}(x)}{M_x}-1,
\qquad x=\lambda b,
\label{eq:saddle_relative_error}
\end{equation}
where $M_{\mathrm{sad}}(x)=\frac{e^x}{r}\Pi_{n_x}
\exp[-D(x,n_x)]$ is the expression on the
right-hand side of Eq.~\eqref{eq:moving_saddle_MFPT}.

Fig.~\ref{fig:moving_saddle_accuracy} shows that
this error decreases toward zero over the threshold
range examined, with pronounced differences among
waiting-time families at finite thresholds.
The decrease is particularly slow for $\kappa=0.25$,
with relative errors at $x=10^5$ of approximately
$-4.52\%$, $-12.21\%$, and $-4.06\%$ for Gamma,
Weibull, and generalized-Gamma ($p=2$) waiting times,
respectively. The moving-saddle formula thus retains
the asymptotic relative-accuracy behavior predicted by
Eq.~\eqref{eq:moving_saddle_MFPT} even in the
subcritical range, where the fixed-index product
develops a growing logarithmic discrepancy.
The family-dependent finite-threshold errors illustrate
the role of the complete renewal law beyond its
short-time exponent.

\section{Discussion}
\label{sec:discussion}
In summary, we have derived an exact series
representation for the MFPT of renewal shot noise with exponentially distributed marks, valid for any finite threshold. Writing the exact finite-threshold MFPT in terms of an auxiliary Poisson average clarifies its relation to fixed-index high-threshold products, including Eq.~(1)
of Ref.~\cite{Bremont2026}.
These approximations evaluate the renewal product at a single index. Although
the auxiliary Poisson index concentrates around its mean as \(x\to\infty\), this concentration alone does not control the renewal-product variation across its
fluctuation window. Therefore, index concentration alone does not guarantee asymptotic agreement in ratio. A crossover occurs when the renewal-induced saddle displacement becomes comparable to the Poisson fluctuation width. Within the short-gap power-law class, the fixed-index
approximation has vanishing relative error for
\(\kappa>1/2\), requires a finite multiplicative correction
at \(\kappa=1/2\), and exhibits a growing discrepancy
on the logarithmic scale for \(0<\kappa<1/2\). 
Our moving-saddle equivalent accounts for this
competition without requiring a waiting-time density or a short-gap power law.

At finite thresholds, however, the MFPT depends on the full renewal law.
Indeed, the exact series samples \(q(j)\) at all positive
integers, making the MFPT sensitive to the full waiting-time
law even at fixed mean waiting time and short-gap exponent.
Consequently, waiting-time laws within the same asymptotic accuracy class can exhibit different finite-threshold corrections and convergence rates.
Comparisons across Gamma, Weibull, and generalized-Gamma renewal laws illustrate this dependence on the waiting-time distribution beyond its short-gap exponent. The empty-state MFPT is also related exactly to the
reciprocal stationary upcrossing rate by the factor \(K_x\).
The limit \(K_x\to1\) recovers the high-threshold
equivalence of these two timescales.

Turning from the mean to the full first-passage distribution, we find a common normalized limit across renewal laws.
For independent exponential marks and a signal started from zero at the beginning of a fresh renewal interval,
\(T_b/\E T_b\) converges in distribution to a unit-mean exponential law as \(b\to\infty\) for any strictly
positive, finite-mean renewal waiting-time distribution. When time is measured in units of the corresponding MFPT,
however, their survival probabilities approach the same
function $
\Pp\!\left(T_b>s\,\E T_b\right)\longrightarrow e^{-s},\ s\ge0$ as $b\to\infty$. Thus, waiting-time laws with different MFPTs approach
the same survival curve when time is measured in units
of the corresponding mean.
This limit can coexist with non-Markovian signal dynamics.
Although successive intervals are independent,
nonexponential waiting times generally give a conditional
residual waiting-time distribution that depends on the
renewal age, which the signal amplitude alone need not
specify.
The role of memory in first-passage kinetics has also
been examined for confined non-Markovian random walks
\cite{Guerin2016}, rare polymer conformational changes
\cite{Levernier2020}, and processes with long-term
correlations \cite{BarbierChebbah2024}.

The exact Poisson-index representation relies on exponential marks, while the explicit crossover classification additionally assumes a short-time
power law for the waiting-time density. For other mark distributions, neither the same averaging kernel nor the same correction factors can be assumed. Whether an analogous smoothing mechanism exists, and which fluctuation scales would govern
its accuracy, remain open questions.

\begin{acknowledgments}
This work was supported by the National Natural Science Foundation of China (Grant No.~12171494) and the Guangdong Basic and Applied Basic Research Foundation
(Grant Nos.~2026A1515011801 and 2026A1515040001).
\end{acknowledgments}

\section*{Data Availability Statement}
The numerical data supporting the findings of this study were generated through theoretical calculations and Monte Carlo simulations. The source code of the simulations will be made publicly available on Zenodo.

\appendix
\section{Exact finite-threshold law from embedded dynamics}
\label{app:exact}
In this appendix, we prove the exact finite-threshold MFPT formula in Eq.~\eqref{eq:exact_MFPT} 
using the embedded chain $R_n=\lambda X(t_n^+)$.
At an arrival time, right continuity and exponential relaxation give
\begin{align}
 R_n&:=
 \lambda\sum_{j=1}^n A_j e^{-\gamma(t_n-t_j)}\notag\\
 &=e^{-\gamma(t_n-t_{n-1})}
   \lambda\sum_{j=1}^{n-1}A_je^{-\gamma(t_{n-1}-t_j)}+\lambda A_n\notag\\
 &=Q_nR_{n-1}+E_n,
 \label{eq:app-embedded-derivation}
\end{align}
where \(Q_n=e^{-\gamma W_n}\in(0,1)\) and
\(E_n\sim\Exp(1)\) are independent. Starting from Eq.~\eqref{eq:app-embedded-derivation}, we derive the exact finite-threshold MFPT through first-step conditioning, a nonlocal initial-value problem,
and a conversion from the arrival count to
physical time. Throughout, we assume only
strictly positive, i.i.d.\ waiting times with
finite mean, without requiring their
distribution to admit a density.

\subsection{Geometric bound and first-step conditioning}

For a deterministic initial level \(R_0=y\in[0,x)\), let
\(\nu_x=\inf\{n\ge1:R_n>x\}\) and
\(u_x(y)=\E_y\nu_x\), where \(\Pp_y\) and
\(\E_y\) denote probability and expectation,
respectively, for the embedded chain
started from \(R_0=y\).
Since \(R_n\ge E_n\), we have
\(\nu_x\le G_x\) almost surely, where
\(G_x:=\inf\{n\ge1:E_n>x\}\).
Under \(\Pp_y\), the rescaled marks \(E_n\) are i.i.d.\
\(\Exp(1)\), so the distribution of
\(G_x\) does not depend on \(y\).
It follows that
\begin{equation}
\Pp_y(\nu_x>k)\le\Pp_y(G_x>k)
=(1-e^{-x})^k,\quad k\ge0.
\label{eq:geom_bound}
\end{equation}
Summing the tail probabilities yields
\begin{equation}
u_x(y)=\sum_{k\ge0}\Pp_y(\nu_x>k)\le e^x,
\end{equation}
where $y\in [0,x)$.
This bound establishes the almost-sure finiteness
of \(\nu_x\) and the boundedness of \(u_x\)
on \([0,x)\). Measurability follows from the
tail-sum representation and the measurable
transition recursion.

To derive the backward equation satisfied by
\(u_x(y)\), we condition on the first
post-arrival state \(R_1=Q_1y+E_1\).
The first arrival contributes one to
\(\nu_x\). If \(R_1>x\), the crossing index equals one.
On \(\{R_1<x\}\), the remaining input pairs
\(\{(Q_{k+1},E_{k+1})\}_{k\ge1}\) are
independent of \(R_1\) and have the same
joint law as the original input sequence.
After shifting the arrival index by one,
the remaining chain therefore has the
conditional law of the original embedded
chain started from \(R_1\). In particular,
\(\E_y[\nu_x-1\mid R_1]=u_x(R_1)\)
on this event.

Conditional on \(Q_1\), the state \(R_1\)
has a shifted exponential distribution,
so \(\Pp_y(R_1=x)=0\).
Taking expectations over \(R_1\) gives
\begin{equation}
 u_x(y)=1+\E_y\!\left[
 u_x(R_1)\ind_{\{R_1<x\}}\right].
 \label{eq:app-first-step}
\end{equation}
In this expectation, \(u_x\) is understood
to be extended by zero outside \([0,x)\).
The indicator restricts the remaining
expectation to subthreshold states, where
\(u_x\) is defined.

For fixed \(q\in(0,1)\), independence of
\(Q_1\) and the unit-rate exponential mark
allows us to integrate over the first mark
\begin{equation}
\begin{aligned}
&\E_y\!\left[u_x(R_1)\ind_{\{R_1<x\}}
\,\middle|\,Q_1=q\right]\\
&\quad=\int_0^{x-qy}u_x(qy+v)e^{-v}\,\dd v\\
&\quad=\int_{qy}^{x}u_x(z)e^{-(z-qy)}\,\dd z,
\end{aligned}
\label{eq:app-transition-integral}
\end{equation}
where the second equality follows from
the substitution \(z=qy+v\).
Both integrals are finite by the preceding
geometric bound.
Taking expectation over \(Q_1\) in
Eq.~\eqref{eq:app-transition-integral}
yields the backward equation
\begin{equation}
 u_x(y)=1+\E_Q\!\left[
 \int_{Qy}^{x}e^{-(z-Qy)}u_x(z)\,\dd z\right],
 \label{eq:backward_main}
\end{equation}
where $y\in[0,x)$ and $\E_Q$ denotes the expectation with respect to $Q$.

\subsection{Nonlocal equation and exact series}

To reduce the backward equation to a
nonlocal initial-value problem, we introduce
\begin{equation}
\begin{aligned}
 H(y)&:=\int_y^x e^{-(z-y)}u_x(z)\,\dd z,\\
 V(y)&:=H(0)-H(y),
\end{aligned}
\label{eq:HV_def}
\end{equation}
where $0\le y\le x$.
Since \(u_x\) is bounded, the representation
\(H(y)=e^y\int_y^x e^{-z}u_x(z)\,\dd z\)
shows that \(H\) is continuous on \([0,x]\),
with \(H(x)=0\).
The backward equation can now be written as
\(u_x(y)=1+\E_Q H(Qy)\).
By dominated convergence, \(u_x\) is
continuous on \([0,x)\), which justifies
differentiation of \(H\) and gives
\(H'(y)=H(y)-u_x(y)\).
Using the definition of \(V(y)\), we obtain
\begin{equation}
\begin{aligned}
 V'(y)
 &=u_x(y)-H(y)\\
 &=1+V(y)-\E_Q V(Qy),\quad V(0)=0.
\end{aligned}
\label{eq:nonlocal_ivp_app}
\end{equation}
Then, we solve Eq.~\eqref{eq:nonlocal_ivp_app} by constructing
a power-series solution.
Recall that \(q(n)=\E Q^n\) and
\(\Pi_n=\prod_{j=1}^n[1-q(j)]\), with \(\Pi_0=1\).
Consider
\begin{equation}
 V(y)=\sum_{n=1}^{\infty}\frac{y^n}{n!}\Pi_{n-1}.
 \label{eq:V_series_app}
\end{equation}
Since \(0<\Pi_n\le1\), on the interval \(0\le y\le x\), the series is dominated by
the exponential series $\sum_{n=1}^{\infty}\frac{x^n}{n!},$ and hence converges uniformly on \([0,x]\)
by the Weierstrass \(M\)-test. The same argument
applies to the derivative series, which is bounded by $\sum_{n=1}^{\infty}\frac{n x^{n-1}}{n!}
=e^x.$ Therefore, termwise differentiation is justified.
Moreover, since \(0\le Q\le1\), we have
\(0\le Qy\le y\), and the same uniform bound
allows the expectation to be interchanged with the
series by dominated convergence. Hence
\[
\E_Q V(Qy)
=
\sum_{n=1}^{\infty}
\frac{y^n}{n!}\Pi_{n-1}\E Q^n
=
\sum_{n=1}^{\infty}
\frac{y^n}{n!}\Pi_{n-1}q(n).
\]
We may therefore differentiate term by term
and interchange expectation with summation,
obtaining $V'(y)=1+\sum_{n=1}^{\infty}\frac{y^n}{n!}\Pi_n$, and
\begin{equation}
\begin{aligned}
 V(y)-\E_QV(Qy)
 &=\sum_{n=1}^{\infty}\frac{y^n}{n!}
 \Pi_{n-1}[1-q(n)]\\
 &=\sum_{n=1}^{\infty}\frac{y^n}{n!}\Pi_n.
\end{aligned}
\label{eq:app-series-verification}
\end{equation}
These expressions show that the series
representation in Eq.~\eqref{eq:V_series_app}
is a solution of the nonlocal equation
\eqref{eq:nonlocal_ivp_app}
with \(V(0)=0\).

To complete the characterization of the
solution, we next prove uniqueness for the
nonlocal equation in Eq.~\eqref{eq:nonlocal_ivp_app}.
Integrating Eq.~\eqref{eq:nonlocal_ivp_app}
gives the equivalent integral equation
\begin{equation}
 V(y)=y+\int_0^y
 \left[V(t)-\E_QV(Qt)\right]\,\dd t.
 \label{eq:V_integral}
\end{equation}
Suppose that \(V\) and \(\widetilde V\) both satisfy
Eq.~\eqref{eq:V_integral}. Let \(D(y):=V(y)-\widetilde V(y)\) denote their difference and define
\(M_D(y)=\max_{0\le z\le y}|D(z)|\).
Since \(0<Q<1\), the integral equation
implies
\(M_D(y)\le2\int_0^y M_D(t)\,\dd t\).
Setting \(C_D:=M_D(x)\) and iterating
this inequality yields
\(M_D(y)\le C_D(2y)^k/k!\)
for every integer \(k\ge1\).
Letting \(k\to\infty\) gives \(D\equiv0\),
which proves uniqueness.

Finally, \(H(x)=0\) implies
\(H(0)=V(x)\), and hence
\(u_x(0)=1+V(x)\).
Substituting the series solution gives
the exact mean number of arrivals
before the first crossing
\begin{equation}
 \E_0\nu_x
 =1+\sum_{n=1}^{\infty}\frac{x^n}{n!}\Pi_{n-1}.
 \label{eq:nu_exact_app}
\end{equation}

\subsection{Conversion to physical time and other initial states}
For a physical initial level \(z\in[0,b)\) with
rescaled level \(y=\lambda z\), the signal
decreases between arrivals, so the first
threshold crossing occurs at the arrival epoch
\(t_{\nu_x}\), and hence
\begin{equation}\label{eq:relation_of_T_b_and_t}
T_b^{(z)}=t_{\nu_x}=\sum_{j=1}^{\nu_x}W_j.
\end{equation}
Let
\(\mathcal F_{j-1}=\sigma((W_k,E_k):1\le k<j)\).
More precisely,
\(\{\nu_x\ge j\}=\bigcap_{k=1}^{j-1}\{R_k\le x\}\),
with the empty intersection interpreted as the whole
sample space. This event depends only on the first
\(j-1\) innovation pairs and is therefore
\(\mathcal F_{j-1}\)-measurable.
Using the independence of \(W_j\) from
\(\mathcal F_{j-1}\), we obtain
\begin{equation}
\begin{aligned}
\E T_b^{(z)}
&=\sum_{j\ge1}
\E_y\!\left[
W_j\ind_{\{\nu_x\ge j\}}
\right]\\
&=m\sum_{j\ge1}\Pp_y(\nu_x\ge j)
=m\E_y\nu_x .
\end{aligned}
\label{eq:app-stopped-sum}
\end{equation}
Here \(m=\E W=r^{-1}\). The expectation \(\E_y\) refers
to the law of the innovations for the embedded chain initialized
at \(R_0=y=\lambda z\), while \(T_b^{(z)}\)
denotes the physical first-passage time from \(X(0)=z\) at the beginning of a
fresh renewal interval.
Combining Eq.~\eqref{eq:app-stopped-sum}
with Eq.~\eqref{eq:nu_exact_app}
for the empty initial state \(z=0\)
establishes Eq.~\eqref{eq:exact_MFPT}.
Retaining the first two terms of the
series gives \(\E T_b\ge(1+x)/r\),
whereas replacing \(\Pi_{n-1}\) by its
upper bound \(1\) yields
\(\E T_b\le e^x/r\).

Furthermore, the exact series also admits an auxiliary Poisson representation.
To obtain this representation, set
\(B_0=1\) and
\(B_n=\Pi_{n-1}\) for \(n\ge1\). 
For an auxiliary $N_x\sim\operatorname{Pois}(x)$,
the Poisson probability mass function gives
\begin{equation}\label{eq:expectation_B_nx}
 \E B_{N_x}
 =e^{-x}\left[1+\sum_{n=1}^{\infty}
 \frac{x^n}{n!}\Pi_{n-1}\right].
\end{equation}
Multiplying Eq.~\eqref{eq:expectation_B_nx} by \(e^x/r\) recovers
the exact identity in
Eq.~\eqref{eq:Poisson_smoothing}.

We next use the same backward equation to
characterize the effect of a nonzero initial
level. For \(0\le y<x\), we have
\(u_x(y)=1+V(x)-\E_QV(Qy)\).
Substituting Eq.~\eqref{eq:V_series_app}
and exchanging expectation and summation
yields
\begin{equation}
 \E_y\nu_x
 =1+\sum_{n=1}^{\infty}\frac{\Pi_{n-1}}{n!}
 \left[x^n-q(n)y^n\right].
 \label{eq:general_initial_nu}
\end{equation}
For a physical initial signal level
\(z=y/\lambda<b\),
the stopped-sum identity
\eqref{eq:app-stopped-sum}
then gives Eq.~\eqref{eq:initial-mean}.
Subtracting this result from the
empty-state MFPT yields
Eq.~\eqref{eq:initial-offset}.
Using \(0<q(n)\Pi_{n-1}\le1\),
we further obtain
\begin{equation}
0\le \E T_b^{(0)}-\E T_b^{(z)}
\le \frac{e^{\lambda z}-1}{r}.
\end{equation}
For fixed \(z\), the upper bound is
independent of \(b\), while
\(\E T_b^{(0)}\to\infty\).
The initial-state correction is therefore
asymptotically negligible, and
\(\E T_b^{(z)}/\E T_b^{(0)}\to1\).

\section{Stationary representation of high-threshold rarity}
\label{app:stationary_rep}

\subsection{Existence and uniqueness of the stationary law}

Let \(\rho=\E Q=q(1)\in(0,1)\). Using
i.i.d.\ copies \((\widehat Q_j,\widehat E_j)\)
of \((Q,E)\), define the perpetuity
\begin{equation}
R
:=\sum_{k=0}^{\infty}
\widehat E_{k+1}\prod_{j=1}^{k}\widehat Q_j.
\label{eq:S_perpetuity}
\end{equation}
Taking expectations of the partial sums and
applying the monotone convergence theorem
gives \(\E R=(1-\rho)^{-1}<\infty\).
Hence the series is finite almost surely.
Separating the first term of the perpetuity
yields the distributional fixed-point relation
\begin{equation}
R\overset d=E+QR',
\label{eq:stationary_fixedpoint}
\end{equation}
where \(R'\overset d=R\) is independent
of the pair \((Q,E)\).
Let \(\pi=\mathcal L(R)\) denote the
probability distribution of \(R\).
We first verify that \(\pi\) is invariant
under the transition of the post-arrival
embedded chain. If the embedded
chain is initialized with \(R_0^\star\sim\pi\),
independently of the future input pairs, then
Eq.~\eqref{eq:stationary_fixedpoint} gives the first update
\(R_1^\star=Q_1R_0^\star+E_1\). For any $z>0$, conditioning on
\(R_0^\star\) gives
\begin{equation}
\begin{aligned}
\Pp(R_1^\star\le z)
&=\int_0^\infty
\Pp(R_1^\star\le z\mid R_0^\star=y)
\,\pi(\dd y)\\
&=\int_0^\infty
\Pp(Q_1y+E_1\le z)\,\pi(\dd y)\\
&=\Pp(QR'+E\le z)\\
&=\Pp(R\le z)=\pi([0,z]),
\end{aligned}
\end{equation}
where the last equality follows from
Eq.~\eqref{eq:stationary_fixedpoint}.
Thus, \(R_1^\star\sim\pi\), showing that
one transition preserves the distribution
\(\pi\).
Since each new input pair is independent
of the current state, the same argument
applies inductively, yielding
\(R_n^\star\sim\pi\) for every \(n\ge0\).
Hence \(\pi\) is a stationary distribution
of the embedded chain.

We now show that this stationary distribution
is unique. Couple two chains with initial states
\(U\) and \(V\) by driving them with the same
future input pairs $\{Q_n,E_n\}_{n\ge1}$. Their difference satisfies
\begin{equation}
R_n^{(U)}-R_n^{(V)}
=\left(\prod_{j=1}^{n}Q_j\right)(U-V).
\label{eq:coupling_diff}
\end{equation}
Since \(\log Q_j=-\gamma W_j\) and the
waiting times are i.i.d.\ with finite mean,
the strong law of large numbers gives $
\frac{1}{n}\sum_{j=1}^{n}\log Q_j
\longrightarrow-\gamma\E W<0
$ almost surely. Consequently,
\(\prod_{j=1}^{n}Q_j\to0\)
almost surely. 

Let \(\widetilde\pi\) be any other stationary
distribution of the embedded chain.
We choose initial states \(U\sim\pi\) and
\(V\sim\widetilde\pi\), with the pair
\((U,V)\) independent of the same future
innovation sequence $\{Q_n,E_n\}_{n\ge 1}$.
Since \(\prod_{j=1}^{n}Q_j\to0\)
almost surely and both initial states
are finite almost surely,
Eq.~\eqref{eq:coupling_diff} gives
\(R_n^{(U)}-R_n^{(V)}\to0\)
almost surely. Let \(g\) be any bounded Lipschitz function.
Since \(R_n^{(U)}-R_n^{(V)}\to0\)
almost surely, we have $
|g(R_n^{(U)})-g(R_n^{(V)})|
\longrightarrow0$
almost surely.
Moreover, this difference is bounded by
\(2\|g\|_\infty\). The dominated convergence
theorem therefore gives
\begin{equation}
\lim_{n\to\infty}
\left[
\E g(R_n^{(U)})
-\E g(R_n^{(V)})
\right]=0.
\end{equation}
On the other hand, since \(\pi\) and
\(\widetilde\pi\) are stationary distributions,
\(R_n^{(U)}\sim\pi\) and
\(R_n^{(V)}\sim\widetilde\pi\) for every
\(n\ge0\). Consequently,
\begin{equation}
\begin{aligned}
&\E g(R_n^{(U)})-\E g(R_n^{(V)})\\
&\quad=
\int_0^\infty g(y)\,\pi(\dd y)
-\int_0^\infty g(y)\,\widetilde\pi(\dd y).
\end{aligned}
\end{equation}
The right-hand side is independent of \(n\),
while the left-hand side tends to zero.
It follows that
\begin{equation}
\int_0^\infty g(y)\,\pi(\dd y)
=
\int_0^\infty g(y)\,\widetilde\pi(\dd y)
\end{equation}
for every bounded Lipschitz function \(g\).
Since bounded Lipschitz functions determine
Borel probability measures on \([0,\infty)\),
we conclude that
\(\pi=\widetilde\pi\), which proves uniqueness of the stationary distribution.

\subsection{Exact stationary moments}
Renewal shot-noise moments have previously been developed using Feynman--Kac and martingale formulations \cite{JangDassiosZhao2018}, while the exponential-mark product moments also appear in the recent high-threshold analysis of Ref.~\cite{Bremont2026}. For completeness, we first justify finiteness
of the moments used in the stationary recursion.
Let
\(R^{(N)}:=\sum_{k=0}^{N-1}
\widehat E_{k+1}\prod_{j=1}^{k}\widehat Q_j\)
be the first \(N\) terms of the nonnegative perpetuity
in Eq.~\eqref{eq:S_perpetuity}, with \(R^{(0)}=0\),
and set \(m_{n,N}=\E[(R^{(N)})^n]\).
These moments are finite because \(0<\widehat Q_j<1\)
and each partial sum is bounded by a finite sum of
exponential random variables. Splitting off the first term
gives \(R^{(N)}\overset d=E+Q\widetilde R^{(N-1)}\),
where \(\widetilde R^{(N-1)}\overset d=R^{(N-1)}\)
is independent of \((Q,E)\).
Expanding the \(n\)th moment gives, with \(q(0)=1\),
\begin{equation}
 m_{n,N}=n!\sum_{j=0}^n
 \frac{q(j)}{j!}m_{j,N-1}.
\end{equation}
Moreover, the identity $
\frac{q(j)}{\Pi_j}
=\frac{1}{\Pi_j}-\frac{1}{\Pi_{j-1}},
j\ge1$
gives $
\sum_{j=0}^{n}\frac{q(j)}{\Pi_j}
=\frac{1}{\Pi_n},$
where \(\Pi_0=q(0)=1\). Here \(m_{0,N}=1\) for all
\(N\ge0\), and \(m_{n,0}=0\) for \(n\ge1\).
We now prove the uniform bound by induction on \(N\),
simultaneously for every integer \(n\ge1\).
The assertion holds at \(N=0\). Assuming
\(m_{j,N-1}\le j!/\Pi_j\) for all \(j\ge1\),
the recursion and telescoping identity yield
\begin{equation}
\begin{aligned}
m_{n,N}
&\le n!\sum_{j=0}^{n}\frac{q(j)}{\Pi_j}=\frac{n!}{\Pi_n}.
\end{aligned}
\end{equation}
Consequently, \(m_{n,N}\le n!/\Pi_n\) for all
integers \(n\ge1\) and \(N\ge0\).
Since \(R^{(N)}\uparrow R\) almost surely,
the monotone convergence theorem yields
\begin{equation}
m_n:=\E R^n
=\lim_{N\to\infty}m_{n,N}
\le\frac{n!}{\Pi_n}<\infty.
\end{equation}
Therefore, we expand the stationary fixed-point
equation and obtain
\begin{equation}
(1-q(n))m_n
=n!\left[1+\sum_{j=1}^{n-1}\frac{q(j)}{j!}m_j\right]=\frac{n!}{\Pi_{n-1}}.
\end{equation}
Thus, we obtain the exact formula for
the stationary moments
\begin{equation}
\E R^n=\frac{n!}{\Pi_n},
\quad n\ge1.
\label{eq:stationary_moments_app}
\end{equation}
Since \(q(n)\to0\), we have \(\log[1-q(n)]\to0\).
The Ces\`aro mean of these logarithms therefore gives
\(n^{-1}\log\Pi_n\to0\), and hence
\(\Pi_n^{-1/n}\to1\). Therefore, the root test
gives
\begin{equation}
\E e^{tR}<\infty,
\qquad 0\le t<1.
\label{eq:R_mgf}
\end{equation}

The exact moment formula 
Eq.~\eqref{eq:stationary_moments_app} will be used
to determine the normalization constant
in the stationary density asymptotics
of Appendix~\ref{app:saddle} and the exponential-moment bound Eq.~\eqref{eq:R_mgf} also provides
tail estimates for the upcrossing analysis
in Appendix~\ref{app:Kx}.

\subsection{Stationary density and upcrossing probability}
We next derive the stationary density
and relate it to the probability of
an upcrossing at a renewal epoch.
Let \(R'\sim\pi\) be independent of
the pair \((Q,E)\), and set \(Z=QR'\).
The stationary fixed-point relation
in Eq.~\eqref{eq:stationary_fixedpoint}
then gives \(R\overset d=Z+E\),
where \(E\sim\Exp(1)\) is independent
of \(Z\).

For any Borel set \(B\subseteq[0,\infty)\),
conditioning on \(Z\) and integrating
over the exponential mark yields
\begin{equation}
\begin{aligned}
\Pp(R\in B)
&=\E\!\left[
  \int_0^\infty
  \ind_{\{Z+e\in B\}}e^{-e}\,\dd e
  \right]\\
&=\E\!\left[
  \int_Z^\infty
  \ind_{\{x\in B\}}e^{-(x-Z)}\,\dd x
  \right]\\
&=\int_B
  \E\!\left[
  e^{-(x-Z)}\ind_{\{Z\le x\}}
  \right]\dd x,
\end{aligned}
\end{equation}
where the second equality follows
from the substitution \(x=Z+e\),
and the last follows from Tonelli's
theorem, since the integrand is
nonnegative. This representation shows that the
stationary law \(\pi\) is absolutely
continuous with respect to Lebesgue
measure. Its density is given by
\begin{equation}
f_R(x)=\E\!\left[
e^{-(x-Z)}\ind_{\{Z\le x\}}
\right],
\qquad x>0.
\end{equation}

It remains to establish continuity and
strict positivity of the stationary density
\(f_R\). We begin with continuity.
Since \(R'\) has an absolutely continuous
distribution, \(Q>0\) almost surely, and \(Q\) is independent of \(R'\),
the product \(Z=QR'\) has no atoms.
Hence \(\Pp(Z=x)=0\) for every \(x>0\).
For each realization of \(Z\), the integrand
\(e^{-(x-Z)}\ind_{\{Z\le x\}}\)
is bounded by one and continuous in \(x\)
except at \(x=Z\). Therefore, 
the dominated convergence theorem
implies that \(f_R\) is continuous on
\((0,\infty)\).

To prove strict positivity of \(f_R\),
we first show that \(\Pp(Z<x)>0\) for every
\(x>0\).
Let \(M=\E R=(1-\rho)^{-1}\)
and choose \(q_0\in(0,1)\) such that
\(\Pp(Q\le q_0)>0\).
Since \(q_0<1\), we may choose an integer
\(k\ge1\) sufficiently large that
\(2Mq_0^k<x/2\).

Consider a stationary chain initialized
with \(R_0^\star\sim\pi\), independently
of the future innovations. We define the event
$
A_k:=
\{R_0^\star\le2M\}
\cap\bigcap_{j=1}^k
\left\{Q_j\le q_0,\,
E_j<\frac{x}{2k}\right\}.$
On \(A_k\), the recursion gives
\begin{equation}
R_k^\star
\le 2Mq_0^k+\frac{x}{2}<x.
\end{equation}
To show that \(A_k\) has positive
probability, we use Markov's inequality to get
\(\Pp(R_0^\star\le2M)\ge1/2\).
Independence of the initial state and
the future innovation pairs gives
\begin{equation}\label{ineq:A_k}
\begin{aligned}
\Pp(A_k)
&=\Pp(R_0^\star\le2M)
\left[
\Pp(Q\le q_0)
\Pp\left(E<\frac{x}{2k}\right)
\right]^k\\
&\ge\frac12
\left[
\Pp(Q\le q_0)
\left(1-e^{-x/(2k)}\right)
\right]^k>0.
\end{aligned}
\end{equation}
Since \(A_k\subseteq\{R_k^\star<x\}\)
and \(R_k^\star\sim\pi\), we obtain
\(\Pp(R<x)\ge\Pp(A_k)>0\) from Eq.~\eqref{ineq:A_k}. Furthermore, \(Z=QR'\le R'\) almost surely,
so
\(\Pp(Z<x)\ge\Pp(R'<x)>0\).
Using the stationary density representation,
we conclude that
\begin{equation}
f_R(x)
=\E\!\left[
e^{-(x-Z)}\ind_{\{Z\le x\}}
\right]
\ge e^{-x}\Pp(Z<x)>0,
\end{equation}
where $x>0$.

Finally, conditioning on the
stationary pre-arrival level \(Z\), we get
\begin{equation}
\begin{aligned}
\Pp(Z\le x<Z+E)
&=\E\!\left[
\ind_{\{Z\le x\}}
\Pp(E>x-Z\mid Z)
\right]\\
&=\E\!\left[
e^{-(x-Z)}\ind_{\{Z\le x\}}
\right]\\
&=f_R(x).
\end{aligned}
\end{equation}
This proves Eq.~\eqref{eq:stationary_density}
and identifies \(f_R(x)\) with the
stationary one-step upcrossing probability
at threshold \(x\).

\subsection{Exponential-functional representation}
To derive the large-threshold asymptotics
of the stationary density \(f_R\), we
represent the stationary level \(R\) as
an exponential functional of an auxiliary
compound-Poisson subordinator in this subsection.
This representation allows us to apply
the density asymptotics of
Haas~\cite{Haas2025} in
Appendix~\ref{app:saddle}, leading to
the moving-saddle asymptotic equivalent
for the MFPT.

Let \(\mathcal N_t\) be an auxiliary
rate-one Poisson process, independent
of i.i.d.\ copies \(\{\widehat W_j\}\)
of \(W\), and define the compound-Poisson subordinator
\begin{equation}
\xi_t:=\gamma\sum_{j=1}^{\mathcal N_t}\widehat W_j.
\label{eq:subordinator}
\end{equation}
Note that the auxiliary Poisson clock in this construction is distinct from the physical renewal arrival process.
The jump times of \(\mathcal N_t\) can be written as
\(0=\sigma_0<\sigma_1<\cdots\). Let
\(H_j=\sigma_j-\sigma_{j-1}\) be the independent
unit-rate exponential holding times.
For the interval \(\sigma_k\le t<\sigma_{k+1}\), we notice that
\(\xi_t=\gamma\sum_{j=1}^{k}\widehat W_j\).
Its exponential functional is defined by
\begin{equation}
I:=\int_0^\infty e^{-\xi_t}\,\dd t.
\label{eq:I_def}
\end{equation}
Thus, for every integer \(m\ge1\), we have
\begin{equation}
\begin{aligned}
\int_0^{\sigma_m}e^{-\xi_t}\,\dd t
&=\sum_{k=0}^{m-1}
\int_{\sigma_k}^{\sigma_{k+1}}
\exp\{-\gamma\sum_{j=1}^k\widehat W_j\}
\,\dd t\\
&=\sum_{k=0}^{m-1}
H_{k+1}\prod_{j=1}^k\widehat Q_j,
\end{aligned}
\end{equation}
where
\(\widehat Q_j=e^{-\gamma\widehat W_j}\)
and the empty product equals one. By the strong law of large numbers,
\(\sigma_m/m\to\E H_1=1\)
almost surely, so
\(\sigma_m\to\infty\).
Since \(e^{-\xi_t}\ge0\),
monotone convergence yields the
pathwise representation
\begin{equation}\label{eq:I_perpetuity}
I=\sum_{k=0}^{\infty}
H_{k+1}\prod_{j=1}^{k}\widehat Q_j,
\end{equation}
where the holding times \(H_j\) have the same
distribution as the exponential marks
\(\widehat E_j\) in
Eq.~\eqref{eq:S_perpetuity} and are
independent of the factors
\(\widehat Q_j\).
Comparing Eqs.~\eqref{eq:S_perpetuity}
and~\eqref{eq:I_perpetuity} shows that
\(R\) and \(I\) have the same probability distribution
\begin{equation}
R\overset d=I.
\label{eq:R_I}
\end{equation}
The Laplace exponent of \(\xi\) is
\begin{equation}
\E e^{-s\xi_t}=\exp[-t\phi(s)],
\label{eq:Laplace_exponent}
\end{equation}
where $\phi(s)=1-q(s)$. 

\section{From stationary upcrossings to first-passage time}
\label{app:Kx}

\subsection{Exponential overshoot and post-crossing dynamics}
We first establish the distribution of the
post-crossing overshoot, which
allows us to represent the subsequent
evolution as a chain with an independent
post-crossing initial state. 

Let
\(\mathcal F_n\) be the sigma-field
generated by the first \(n\) innovation
pairs, and define
\(\mathcal G_n=\mathcal F_{n-1}
\vee\sigma(Q_n)\).
Since
\(\{\nu_x\ge n\}
=\bigcap_{j=1}^{n-1}\{R_j\le x\}\)
is \(\mathcal F_{n-1}\)-measurable,
the event \(\{\nu_x\ge n\}\) is also
\(\mathcal G_n\)-measurable.
On this event, the residual gap $
d_n:=x-Q_nR_{n-1}\ge0$
is \(\mathcal G_n\)-measurable.
Since \(R_n=Q_nR_{n-1}+E_n\),
the first crossing occurs at step \(n\)
precisely when
\(\{\nu_x\ge n\}\cap\{E_n>d_n\}\).

For \(s\ge0\), the additional condition
\(R_n-x>s\) is equivalent to
\(E_n>d_n+s\).
Because \(E_n\sim\Exp(1)\) is independent
of \(\mathcal G_n\), conditioning on
\(\mathcal G_n\) gives
\begin{equation}
\begin{aligned}
&\Pp_0(\nu_x=n,\ R_n-x>s)\\
&=\E_0\!\left[
\ind_{\{\nu_x\ge n\}}
\Pp(E_n>d_n+s\mid\mathcal G_n)
\right]\\
&=\E_0\!\left[
\ind_{\{\nu_x\ge n\}}e^{-(d_n+s)}
\right]\\
&=e^{-s}\E_0\!\left[
\ind_{\{\nu_x\ge n\}}e^{-d_n}
\right]\\
&=e^{-s}\Pp_0(\nu_x=n).
\end{aligned}
\label{eq:app-exponential-overshoot}
\end{equation}
The last equality follows by applying
the same conditional argument with \(s=0\).
Then, we define the overshoot at the first crossing by
\(O_x:=R_{\nu_x}-x\).
Since \(\nu_x<\infty\) almost surely,
summing Eq.~\eqref{eq:app-exponential-overshoot}
over \(n\ge1\) yields
\begin{equation}
\begin{aligned}
\Pp_0(O_x>s)
&=\sum_{n\ge1}
\Pp_0(\nu_x=n,\ O_x>s)\\
&=e^{-s}\sum_{n\ge1}\Pp_0(\nu_x=n)
=e^{-s}.
\end{aligned}
\end{equation}
Moreover,
Eq.~\eqref{eq:app-exponential-overshoot}
shows that the conditional distribution
of \(O_x\) given \(\nu_x=n\) does not
depend on \(n\).

Hence \(O_x\sim\Exp(1)\) and is
independent of the first-crossing
index \(\nu_x\). Finally, since the innovation pairs are
i.i.d.\ and \(\nu_x\) is a stopping time,
the future inputs after \(\nu_x\) are
independent of the history up to that
crossing and retain their original law.
The post-crossing chain can therefore
be represented as an independent chain
started from
\(Y_x:=x+\widetilde E\), where
\(\widetilde E\sim\Exp(1)\) is independent
of the future innovations.

\subsection{Stationary upcrossings and the exact MFPT identity}
We now relate the stationary upcrossing
probability to the mean first-crossing index.
Using the post-crossing law established
above, we derive a renewal relation for
the upcrossing generating functions and
obtain an exact identity involving
the correction factor \(K_x\).
The input innovations after the crossing are
also independent of its past.
Accordingly, the first-upcrossing decomposition below
may restart from \(Y_x\). For any initial state \(U\)
independent of the future innovations, write
$
v_n(U):=\Pp\{Q_nR_{n-1}^{(U)}\le x<R_n^{(U)}\}
=u_n^{(U)}(x).
$
For random \(U\), this probability includes averaging over
its initial distribution. We define
\begin{equation}
K_x
:=
1+\sum_{n=1}^{\infty}[v_n(Y_x)-v_n(0)].
\label{eq:Kx_def}
\end{equation}
Then, we drive the chains from \(Y_x\) and \(0\)
with the same future inputs \(\{Q_n,E_n\}_{n\ge 1}\). Their
post-arrival difference satisfies
\begin{equation}
\Delta_n=Y_x\prod_{j=1}^{n}Q_j,
\quad
\E\Delta_n=(x+1)\rho^n,
\quad \rho=q(1)<1.
\end{equation}
We next control the dependence of the
upcrossing probability on the initial state.
For \(E\sim\Exp(1)\) and deterministic
\(a,b\ge0\), direct comparison of the shifted
exponential densities gives
\begin{equation}
 \|\mathcal L(E+a)-\mathcal L(E+b)\|_{\mathrm{TV}}
 =1-e^{-|a-b|}\le|a-b|.
\end{equation}
Consequently, for every measurable
\(0\le g\le1\),
\(\lvert\E g(E+a)-\E g(E+b)\rvert\le|a-b|\).

Let \(U\) and \(V\) be two nonnegative
initial states, independent of the future
innovation pairs, with
\(\E|U-V|<\infty\).
We couple the corresponding chains
\(R_j^{(U)}\) and \(R_j^{(V)}\)
by using the same sequence
\(\{(Q_j,E_j)\}_{j\ge1}\).
Conditioning on the final exponential
mark and applying the shifted-exponential
total-variation bound, we obtain, for
every measurable \(0\le g\le1\),
\begin{equation}
\begin{aligned}
&|\E g(R_j^{(U)})-\E g(R_j^{(V)})|\\
&\quad\le
\E|R_j^{(U)}-R_j^{(V)}|=\rho^j\E|U-V|.
\end{aligned}
\end{equation}
We apply this contraction estimate to the
upcrossing probabilities. For fixed
\(q\in(0,1)\), we define the bounded test function $g_{x,q}(z)=e^{-(x-qz)}\ind_{\{qz\le x\}}$, which 
is Borel measurable and takes values in \([0,1]\).
Since \(Q_n\) is independent of the
states at step \(n-1\), conditioning on
\(Q_n\) gives
\begin{equation}
v_n(U)
=\E_Q\!\left[
\E g_{x,Q}(R_{n-1}^{(U)})
\right],
\end{equation}
and the same representation holds for
\(v_n(V)\).
For each fixed \(q\), the test-function
contraction at step \(n-1\) gives
\begin{equation}
\left|
\E g_{x,q}(R_{n-1}^{(U)})
-\E g_{x,q}(R_{n-1}^{(V)})
\right|
\le\rho^{n-1}\E|U-V|,
\end{equation}
where the bound is uniform in \(q\in(0,1)\).
Averaging over \(Q_n\), we obtain
\begin{equation}
|v_n(U)-v_n(V)|
\le\rho^{n-1}\E|U-V|,
\ n\ge2.
\label{eq:vn_coupling_bound_general}
\end{equation}
For \(n=1\), both upcrossing
probabilities lie in \([0,1]\), so their
difference is bounded by one.
Consequently, combining with Eq.~\eqref{eq:vn_coupling_bound_general} gives
\begin{equation}
\begin{aligned}
&\sum_{n=1}^{\infty}
|v_n(U)-v_n(V)|\le
1+\frac{\rho}{1-\rho}\E|U-V|
<\infty,
\end{aligned}
\end{equation}
since $\rho\in(0,1)$ and $\E|U-V|<\infty$.
Thus, we apply the preceding estimate established in Eq.~\eqref{eq:vn_coupling_bound_general}, for \(n\ge2\),
\begin{equation}
|v_n(Y_x)-v_n(0)|
\le(x+1)\rho^{n-1},
\label{eq:vn_coupling_bound}
\end{equation}
which ensures absolute convergence of
Eq.~\eqref{eq:Kx_def} for each fixed \(x\).
In addition, coupling the process from \(Y_x\) with an
independent stationary state \(R\) and applying Eq.~\eqref{eq:vn_coupling_bound_general} gives
\begin{equation}\label{eq:v_n_coupling_R_bounding}
 |v_n(Y_x)-f_R(x)|
 \le(x+1+\E R)\rho^{n-1},\ n\ge2,
\end{equation}
which provides absolute convergence of
$\sum_{n=1}^{\infty}
|v_n(Y_x)-v_n(R)|$. 
Here the equality \(v_n(R)=f_R(x)\) holds
at every step because the chain is stationary.
In particular, \(v_n(Y_x)\to f_R(x)\).
For an integer $L\ge 1$, define the partial sum
\begin{equation}
K_{x,L}
=
1+\sum_{n=1}^{L}
\left[v_n(Y_x)-v_n(0)\right].
\label{eq:Kx_partial_sum}
\end{equation}
The absolute convergence established above allows us to express
the truncation error as the remaining tail. Since
$n\ge L+1\ge 2$, the coupling estimate applies to every term
in this tail and gives
\begin{align}
\left|K_x-K_{x,L}\right|
&=
\left|
\sum_{n=L+1}^{\infty}
\left[v_n(Y_x)-v_n(0)\right]
\right|
\nonumber\\
&\le
\sum_{n=L+1}^{\infty}
\left|v_n(Y_x)-v_n(0)\right|
\nonumber\\
&\le
(x+1)\sum_{n=L+1}^{\infty}\rho^{n-1}
\nonumber\\
&=
\frac{(x+1)\rho^L}{1-\rho}
=:\tau_{x,L}.
\label{eq:Kx_tail_bound}
\end{align}
This proves Eq.~\eqref{eq:K-truncation-main} for every $x>0$. For a prescribed absolute tolerance $\varepsilon>0$, the
condition $\tau_{x,L}\le\varepsilon$ is equivalent to
\begin{equation}
L|\log\rho|
\ge
\log\!\left[
\frac{x+1}{\varepsilon(1-\rho)}
\right].
\label{eq:K-cutoff-condition}
\end{equation}
Thus, for any integer $L\ge L_\varepsilon(x)$, where
\begin{equation}
L_\varepsilon(x)
=
\max\left\{
1,\,
\left\lceil
\frac{
\log\!\left[(x+1)/\{\varepsilon(1-\rho)\}\right]
}{
|\log\rho|
}
\right\rceil
\right\},
\label{eq:K-cutoff}
\end{equation}
ensures
$\left|K_x-K_{x,L}\right|\le\varepsilon$.
For fixed $\rho$, this sufficient cutoff grows logarithmically
with $(x+1)/\varepsilon$ as that ratio tends to infinity.

Let $
G_0(z)=\sum_{n\ge1}v_n(0)z^n, G_+(z)=\sum_{n\ge1}v_n(Y_x)z^n$
and \(F(z)=\E_0z^{\nu_x}\).
Write \(h_k=\Pp_0(\nu_x=k)\). The generating functions are finite for \(0\le z<1\), as the
upcrossing probabilities are bounded by one.
For an upcrossing at step \(n\), either the
first crossing occurs there, or its index is
\(k<n\) and a subsequent upcrossing occurs
\(n-k\) steps after the independent restart
from \(Y_x\). The overshoot calculation above
gives the convolution
\begin{equation}\label{eq:convolution_of_v_n}
 v_n(0)=h_n+\sum_{k=1}^{n-1}h_kv_{n-k}(Y_x).
\end{equation}
For Eq.~\eqref{eq:convolution_of_v_n}, 
multiplying the first-upcrossing decomposition
by \(z^n\) and summing over \(n\ge1\),
we obtain
\begin{align}
G_0(z)
&=F(z)+
\sum_{n=1}^{\infty}
\sum_{k=1}^{n-1}
h_k v_{n-k}(Y_x)z^n
\notag\\
&=F(z)+
\sum_{k=1}^{\infty}
\sum_{m=1}^{\infty}
h_k v_m(Y_x)z^{k+m}
\notag\\
&=F(z)\bigl[1+G_+(z)\bigr],
\label{eq:renewal_generating}
\end{align}
where \(0\le z<1\).
The interchange of summation in Eq.~\eqref{eq:renewal_generating} follows
from Tonelli's theorem, since all terms
are nonnegative. 

To determine the behavior of \(G_+(z)\)
as \(z\uparrow1\), we separate the stationary
upcrossing probability from each term
\begin{align}
G_+(z)
&=\sum_{n=1}^{\infty}v_n(Y_x)z^n
\notag\\
&=\sum_{n=1}^{\infty}f_R(x)z^n
+\sum_{n=1}^{\infty}
[v_n(Y_x)-f_R(x)]z^n
\notag\\
&=f_R(x)\frac{z}{1-z}
+\sum_{n=1}^{\infty}
[v_n(Y_x)-f_R(x)]z^n.
\end{align}
Applying the stationary coupling estimate in Eq.~\eqref{eq:v_n_coupling_R_bounding} gives $
\sum_{n=1}^{\infty}
|v_n(Y_x)-f_R(x)|<\infty.$
The second term is therefore uniformly
bounded for \(0\le z<1\), and hence
\begin{equation}\label{eq:G_+_bound}
(1-z)G_+(z)\longrightarrow f_R(x),
\qquad z\uparrow1.
\end{equation}

The geometric domination in
Eq.~\eqref{eq:geom_bound} gives
\(\E_0\nu_x<\infty\), and
\begin{equation} \label{eq:limit_of_F}
\frac{1-F(z)}{1-z}
 =\E_0\!\left[\sum_{j=0}^{\nu_x-1}z^j\right]
 \longrightarrow\E_0\nu_x ,\quad z\uparrow1.
\end{equation}
Moreover, the generating-function identity
can be rearranged as $
 [1-F(z)][1+G_+(z)]
 =1+\sum_{n\ge1}[v_n(Y_x)-v_n(0)]z^n$.
Note that the right-hand series converges absolutely
at \(z=1\) by Eq.~\eqref{eq:vn_coupling_bound}.
Taking \(z\uparrow1\) and combining the
limits given in Eqs.~\eqref{eq:G_+_bound} and ~\eqref{eq:limit_of_F}
therefore yields
\begin{equation}
\begin{aligned}
\E_0\nu_x
&=\lim_{z\uparrow1}\frac{1-F(z)}{1-z}
\\
&=\frac{1+\sum_{n\ge1}[v_n(Y_x)-v_n(0)]}{f_R(x)}\\
&=\frac{K_x}{f_R(x)}.
\end{aligned}
\label{eq:nu_Kx}
\end{equation}
Using \(\E T_b=r^{-1}\E\nu_x\) then gives
the exact identity in Eq.~\eqref{eq:MFPT_Kx}.
The exact MFPT identity converts the series remainder into
an error bound for the finite-threshold correction.
Since $T_{\mathrm{RE}}(b)=[r f_R(x)]^{-1}$, we have
\begin{equation}
\left|
\frac{\mathbb{E}T_b}{T_{\mathrm{RE}}(b)}
-K_{x,L}
\right|=
\left|K_x-K_{x,L}\right|
\le
\tau_{x,L}.
\label{eq:MFPT-correction-error-app}
\end{equation}
Subtracting one from the corresponding two-sided inequality
gives Eq.~\eqref{eq:MFPT-correction-interval}.
Hence $K_{x,L}-1$ approximates the relative correction
$\mathbb{E}T_b/T_{\mathrm{RE}}(b)-1$ with an absolute
truncation error at most $\tau_{x,L}$.
If a numerical approximation $\widehat K_{x,L}$ to the retained
sum satisfies
$\left|\widehat K_{x,L}-K_{x,L}\right|\le\eta_L$,
the triangle inequality gives
\begin{equation}
\left|K_x-\widehat K_{x,L}\right|
\le
\tau_{x,L}+\eta_L.
\label{eq:K-total-error}
\end{equation}
Thus, a bound on the error in evaluating the retained terms
can be added directly to the geometric truncation bound.

\subsection{High-threshold limit of the correction factor}
To obtain the large-threshold MFPT
equivalent, we want to show that
\(K_x\to1\) as \(x\to\infty\).
Let us split the defining series at a
logarithmic cutoff and estimate the
early- and late-arrival contributions
separately. By choosing
$
L_x:=\left\lceil\frac{3\log x}{|\log\rho|}\right\rceil,\ 
\varepsilon_x:=x^{-1/2}$ and combining Eq.~\eqref{eq:Kx_tail_bound}, 
we have
\begin{equation}
|K_x-K_{x,L_x}|=O(x^{-2}), \ x\to\infty.
\label{eq:Kx_tail_o}
\end{equation}

For the chain started from zero,
coupling it with a stationary chain driven
by the same innovations gives
\(R_n^{(0)}\le R_n^\star\) almost surely,
where \(R_n^\star\overset d=R\).
Since an upcrossing at step \(n\)
requires \(R_n^{(0)}>x\), we have
\begin{equation}
\begin{aligned}
\sum_{n=1}^{L_x}v_n(0)
&\le\sum_{n=1}^{L_x}
\Pp(R_n^{(0)}>x)\\
&\le L_x\Pp(R>x)\\
&\le L_x\E e^{tR}e^{-tx}
\longrightarrow0,
\end{aligned}
\label{eq:low_state_early}
\end{equation}
for any fixed \(t\in(0,1)\) as $x\to\infty$, where
the last limit follows from
Eq.~\eqref{eq:R_mgf} and
\(L_x=O(\log x)\).

For the chain started from \(Y_x\), we set
\(R_n^{(+)}:=R_n^{(Y_x)}\) and introduce the
cumulative relaxation exponent
\(S_n:=\gamma\sum_{j=1}^{n}W_j\).
Since $Y_x\overset d= x+\widetilde E$, the synchronous coupling gives
\begin{equation}
R_n^{(+)}=e^{-S_n}(x+\widetilde E)+R_n^{(0)}.
\end{equation}
Let
$p_x:=\Pp(\gamma W<\varepsilon_x)\longrightarrow0
$ as $x\to\infty$,
because \(W>0\) almost surely. If
\(S_n<\varepsilon_x\), each of the first
\(n\) waiting times satisfies
\(\gamma W_j<\varepsilon_x\); independence yields
\begin{equation}
\Pp(S_n<\varepsilon_x)\le p_x^n.
\end{equation}
On \(S_n\ge\varepsilon_x\), for sufficiently large \(x\),
\begin{equation}
x-e^{-S_n}x
\ge x(1-e^{-\varepsilon_x})
\ge c\sqrt{x}
\end{equation}
for a fixed \(c>0\). Thus, on this event,
\(R_n^{(+)}>x\) requires a contribution of
order \(\sqrt{x}\) from \(\widetilde E\) or
\(R_n^{(0)}\). For any fixed \(t\in(0,1)\),
\(\E e^{tR_n^{(0)}}\le\E e^{tR}<\infty\)
by monotone coupling from the zero state.
Combining the exponential tail of
\(\widetilde E\) with Markov's inequality
for \(e^{tR_n^{(0)}}\) therefore yields
constants \(C,c_0>0\), independent of
\(n\le L_x\), such that
\begin{equation}
\Pp(R_n^{(+)}>x)
\le
p_x^n+Ce^{-c_0\sqrt{x}}.
\label{eq:Rplus_bound}
\end{equation}
Consequently,
\begin{equation}
\sum_{n\le L_x}v_n(Y_x)
\le
\frac{p_x}{1-p_x}
+CL_xe^{-c_0\sqrt{x}}
\longrightarrow0,\ x\to\infty.
\label{eq:high_state_early}
\end{equation}
The tail estimate \eqref{eq:Kx_tail_o} and
the two early-time estimates
\eqref{eq:low_state_early} and
\eqref{eq:high_state_early} now give
\(K_x\to1\). Combining this limit with
Eq.~\eqref{eq:nu_Kx} and the stopped-sum relation
\(\E T_b=r^{-1}\E_0\nu_x\) establishes the
large-threshold equivalent in Eq.~\eqref{eq:MFPT_Kx}.

\section{Stationary density asymptotics and the moving-saddle equivalent}
\label{app:saddle}
In this appendix, we aim to prove the moving-saddle equivalent in Eq.~\eqref{eq:moving_saddle_MFPT} for general strictly positive waiting times with finite mean.
The derivation proceeds by applying the
stationary density asymptotics of
Haas~\cite{Haas2025}, identifying the
prefactor using the exact stationary moments,
and combining the resulting density
equivalent with Eq.~\eqref{eq:MFPT_Kx}, which is obtained in Appendix~\ref{app:Kx}.
The analysis applies to strictly positive
renewal waiting times with finite mean,
without requiring a waiting-time density.
At the end, we impose the additional short-gap
power-law condition to establish the three
regimes of asymptotic accuracy, as shown in Eqs.~\eqref{eq:regime_super}--\eqref{eq:regime_sub}.
\subsection{General stationary density asymptotics}
We first establish the density equivalent under the
general assumptions \(W>0\) almost surely and
\(m=\E W\in(0,\infty)\). For \(s>0\),
differentiation under the expectation gives
\(q'(s)=-\gamma\E[We^{-\gamma sW}]\) and $
s\phi'(s)
=\mathbb E[\gamma sW e^{-\gamma sW}]$.
Since \(W>0\) almost surely,
both \(e^{-\gamma sW}\) and
\(\gamma sW e^{-\gamma sW}\)
converge to zero almost surely as
\(s\to\infty\).
Moreover,
\begin{equation}\label{eq:upper_bound_of_eW}
0\le e^{-\gamma sW}\le1,
\quad
0\le\gamma sW e^{-\gamma sW}\le e^{-1},
\end{equation}
where the second bound follows from
\(ve^{-v}\le e^{-1}\) for \(v\ge0\).
In Eq.~\eqref{eq:upper_bound_of_eW}, both upper bounds are integrable and
independent of \(s\). Therefore,
the dominated convergence theorem
gives
\begin{equation}
\lim_{s\to\infty}q(s)
=\mathbb E\left[\lim_{s\to\infty}e^{-\gamma sW}\right]
=0,
\end{equation}
and
\begin{equation}\label{eq:limit_of_sphi(s)}
\lim_{s\to\infty}s\phi'(s)
=\mathbb E\left[
\lim_{s\to\infty}\gamma sW e^{-\gamma sW}
\right]=0.
\end{equation}
Then we can get
\begin{equation}\label{eq:Hass_assume}
  \lim_{s\to\infty}\frac{s\phi'(s)}{\phi(s)}=0. 
\end{equation}
Let \(g(s)=s/\phi(s)\).
For \(s>0\), we have
\begin{equation}
\phi(s)-s\phi'(s)
=\E\!\left[
1-(1+\gamma sW)e^{-\gamma sW}
\right]>0,
\end{equation}
and hence
\begin{equation}
g'(s)
=\frac{\phi(s)-s\phi'(s)}{\phi(s)^2}>0.
\label{eq:hprime}
\end{equation}
Thus, \(g\) is continuous and strictly
increasing on \((0,\infty)\). To determine the range of \(g\), note that $
\frac{\phi(s)}{s}
=\E\!\left[
\frac{1-e^{-\gamma sW}}{s}
\right].$
Since
\(0\le(1-e^{-\gamma sW})/s\le\gamma W\)
and \(\E W=m<\infty\), dominated
convergence gives
\begin{equation}
\lim_{s\downarrow0}g(s)
=\frac{1}{\gamma m}
=\frac{r}{\gamma}.
\end{equation}
Moreover, \(q(s)\to0\) as \(s\to\infty\),
so \(\phi(s)\to1\) and \(g(s)\sim s\to\infty\).
Consequently, \(g\) maps \((0,\infty)\)
bijectively onto \((r/\gamma,\infty)\),
and its inverse
\(\chi=g^{-1}\) is well defined
on this interval. Thus, $\chi(x)\to\infty$ when $x\to\infty$.
Differentiating
\(g(\chi(x))=x\) and combining with Eq.~\eqref{eq:limit_of_sphi(s)} yields
\begin{equation}
\chi'(x)=
\frac{\phi(\chi(x))^2}
{\phi(\chi(x))-\chi(x)\phi'(\chi(x))}
\longrightarrow1,\ (x\to\infty).
\label{eq:chi_prime}
\end{equation}
Also, \(\chi(x)/x=\phi(\chi(x))=1-q(\chi(x))\to1\) as $x\to\infty$. 
The exponential-functional representation in
Eq.~\eqref{eq:R_I} allows us to apply the
density asymptotics of Haas~\cite{Haas2025}
(Theorem~1 in the arXiv version). The theorem
concerns the exponential functional of a
drift-free, unkilled subordinator with a nonzero
L\'evy measure, under condition (H), $
\limsup_{s\to\infty}\frac{s\phi'(s)}{\phi(s)}<1.$
Our auxiliary compound-Poisson subordinator in
Eq.~\eqref{eq:subordinator} satisfies these
structural assumptions. Its L\'evy measure is
\(\nu_\xi(B)=\Pp(\gamma W\in B)\) for Borel
sets \(B\subset(0,\infty)\), with
\(\nu_\xi((0,\infty))=1\), and its Laplace
exponent is \(\phi(s)=1-q(s)\).
Since \(q(s)\to0\) and \(s\phi'(s)\to0\),
as established in Eq.~\eqref{eq:Hass_assume},
$\lim_{s\to\infty}\frac{s\phi'(s)}{\phi(s)}=0,
$
which verifies condition (H).

The inverse function \(\psi\) in Haas's theorem,
defined by \(\psi(x)/\phi(\psi(x))=x\),
coincides with our moving-saddle function
\(\chi(x)\). Its domain begins at
\(x_\psi:=\lim_{s\downarrow0}s/\phi(s)=r/\gamma\).
Moreover, \(R\overset d=I\) by
Eq.~\eqref{eq:R_I}, so the density
\(k\) in Haas's theorem is our stationary
density \(f_R\). Consequently, for any fixed
\(x_0>r/\gamma\),
\begin{equation}
\begin{aligned}
f_R(x)&=C\sqrt{\chi'(x)}
\exp\left[-\int_{x_0}^{x}\frac{\chi(u)}{u}\,\dd u\right]\\
&\quad\times\left[1+O\!\left(\frac1{\chi(x)}\right)\right],
\end{aligned}
\label{eq:Haas_density}
\end{equation}
where \(C>0\) depends on the renewal law
and the chosen lower integration limit.

Minchev and Savov~\cite{MinchevSavov2023}
also established leading density asymptotics
under a positive-increase condition. We use
Haas's theorem here for its explicit relative
remainder and the prefactor in terms of the
renewal product is identified independently
through the exact moments below.

\subsection{Moment localization and prefactor identification}
To identify the constant \(C\) in
Eq.~\eqref{eq:Haas_density}, we evaluate
the stationary moments using Laplace's method
and match the resulting asymptotics with
the exact moment formula in
Eq.~\eqref{eq:stationary_moments_app}.
Define
\begin{equation}
J(y):=\int_{x_0}^{y}\frac{\chi(u)}{u}\,\dd u,
\quad
F_n(y):=n\log y-J(y),
\end{equation}
for \(y>x_0\). Since \(J'(y)=\chi(y)/y\),
differentiation gives
\begin{equation}
F_n'(y)=\frac{n-\chi(y)}{y}.
\end{equation}
The stationary-point equation
\(F_n'(y)=0\) is equivalent to
\(\chi(y)=n\).
Since \(\chi=g^{-1}\), its unique
solution is
\begin{equation}
y_n:=g(n)=\frac{n}{\phi(n)},
\qquad
\chi(y_n)=n.
\end{equation}
Moreover, \(\phi(n)\to1\) implies
\(y_n\sim n\to\infty\), so
\(y_n>x_0\) for sufficiently large \(n\).

As \(\chi\) is strictly increasing,
\(F_n'(y)>0\) for \(x_0<y<y_n\)
and \(F_n'(y)<0\) for \(y>y_n\).
Thus, \(y_n\) is the unique global
maximum of \(F_n\) on \((x_0,\infty)\).
Differentiating once more, we obtain
\begin{equation}
F_n''(y)
=-\frac{\chi'(y)}{y}
-\frac{n-\chi(y)}{y^2}.
\end{equation}
Using \(\chi(y_n)=n\) and
\(\chi'(y_n)\to1\) from
Eq.~\eqref{eq:chi_prime}, we find
\begin{equation}
F_n''(y_n)
=-\frac{\chi'(y_n)}{y_n}
\sim-\frac1{y_n}.
\end{equation}
Therefore, the curvature at \(y_n\)
identifies \(\sqrt{y_n}\) as the natural
localization scale around the saddle.
To justify Laplace's method on this scale,
we analyze the central window and establish
bounds for the complementary regions.

Fix \(\varepsilon\in(0,1)\).
Since the relative error in
Eq.~\eqref{eq:Haas_density} tends to zero
and \(\chi'(y)\to1\) as \(y\to\infty\),
we may choose \(Y>x_0\) such that,
for every \(y\ge Y\),
\begin{equation}
\left|
\frac{f_R(y)}
{C\sqrt{\chi'(y)}e^{-J(y)}}-1
\right|\le\varepsilon,
\quad
\frac12\le\chi'(y)\le\frac32.
\end{equation}
The cutoff \(Y\) is fixed independently
of \(n\) in the following estimates. With \(Y\) fixed as above, we decompose
the moment integral as
\begin{equation}
\begin{aligned}
\E R^n
&=\int_0^\infty y^nf_R(y)\,\dd y\\
&=\int_0^Y y^nf_R(y)\,\dd y
+\int_Y^\infty y^nf_R(y)\,\dd y.
\end{aligned}
\end{equation}
Since \(f_R\) is a probability density,
the contribution from the bounded interval
satisfies
\begin{equation}
\int_0^Y y^nf_R(y)\,\dd y\le Y^n.
\end{equation}
Moreover, the identity
\(\chi(u)/u=\phi(\chi(u))<1\) gives
\(J(y_n)\le y_n-x_0\) for sufficiently
large \(n\).
Consequently,
\begin{equation}\label{eq:app-saddle-bounded}
\frac{\displaystyle
\int_0^Y y^nf_R(y)\,\dd y}
{C\sqrt{y_n}e^{F_n(y_n)}}\le
\frac{e^{-x_0}}{C\sqrt{y_n}}
\left(
\frac{Ye^{y_n/n}}{y_n}
\right)^n
\longrightarrow0,
\end{equation}
since \(Y\) is fixed and \(y_n\sim n\).
Thus, the bounded portion is negligible
relative to the saddle scale.

Having established this estimate, we
consider the integral over \([Y,\infty)\).
We first evaluate the contribution from
a central window around \(y_n\), followed
by bounds on the remaining tails.

Fix \(K>0\) and define
\begin{equation}
I_{n,K}:=
\int_{y_n-K\sqrt{y_n}}^{y_n+K\sqrt{y_n}}
y^nf_R(y)\,\dd y.
\end{equation}
Since \(y_n\sim n\to\infty\), the central
window lies in \([Y,\infty)\) for all
sufficiently large \(n\).
Set $y_{n,z}:=y_n+z\sqrt{y_n},$ where $|z|\le K.$
The convergence \(\chi'(y)\to1\)
is uniform on the moving window
\(|y-y_n|\le K\sqrt{y_n}\).
More precisely, let $
\omega_n:=
\sup_{|t|\le K\sqrt{y_n}}
|\chi'(y_n+t)-1|.$
Then \(\omega_n\to0\), and
\(\chi(y_n)=n\) gives $
\chi(y_n+t)-n=t+r_n(t),$
where$
|r_n(t)|\le\omega_n|t|.$ Using \(F_n'(y)=[n-\chi(y)]/y\), we therefore obtain
\begin{equation}
\begin{aligned}
F_n(y_{n,z})-F_n(y_n)
&=-\int_0^{z\sqrt{y_n}}
\frac{\chi(y_n+t)-n}{y_n+t}
\,\dd t\\
&=-\frac{z^2}{2}+o(1),
\end{aligned}
\end{equation}
uniformly for \(|z|\le K\).
To evaluate the central integral, we define
the relative density error by $
\delta(y):=
\frac{f_R(y)}
{C\sqrt{\chi'(y)}e^{-J(y)}}-1.$
Eq.~\eqref{eq:Haas_density} gives
\(\delta(y)\to0\) as \(y\to\infty\).
Since $
\inf_{|z|\le K}y_{n,z}
=y_n-K\sqrt{y_n}\to\infty,$
both
\(\chi'(y_{n,z})\to1\) and
\(\delta(y_{n,z})\to0\) hold uniformly
for \(|z|\le K\). Substituting \(y=y_{n,z}\) into
\(I_{n,K}\) and using the density
representation, we obtain
\begin{equation}\label{eq:limit_of_central}
\begin{aligned}
\frac{I_{n,K}}
{C\sqrt{y_n}e^{F_n(y_n)}}&=\int_{-K}^{K}
\sqrt{\chi'(y_{n,z})}
e^{F_n(y_{n,z})-F_n(y_n)}\\
&\qquad\times
[1+\delta(y_{n,z})]\,\dd z\\
&\longrightarrow
\int_{-K}^{K}e^{-z^2/2}\,\dd z,
\qquad n\to\infty.
\end{aligned}
\end{equation}
The passage to the limit follows
from uniform convergence of the
integrand on the fixed interval
\([-K,K]\).

Having established the Gaussian limit
for the central contribution, we turn
to the complementary regions.
We first derive a Gaussian upper bound
for \(F_n\) on \([y_n/2,2y_n]\),
which will control the intermediate tails.
Since \(y_n\to\infty\) and
\(\chi'(y)\to1\), we have $
\frac12\le\chi'(y)\le\frac32,$ and 
$
\frac{y_n}{2}\le y\le2y_n,
$
for all sufficiently large \(n\).
Using \(\chi(y_n)=n\) and the
mean-value theorem, we obtain $
|\chi(y)-n|
\ge\frac12|y-y_n|,
\frac{y_n}{2}\le y\le2y_n.$
Since \(y\le2y_n\), this gives
\begin{equation}
|F_n'(y)|
=\frac{|n-\chi(y)|}{y}
\ge\frac{|y-y_n|}{4y_n}.
\end{equation}
Moreover, \(F_n'(y)>0\) for \(y<y_n\)
and \(F_n'(y)<0\) for \(y>y_n\).
For \(y_n\le y\le2y_n\),
integration gives
\begin{equation}
\begin{aligned}
F_n(y)-F_n(y_n)&=\int_{y_n}^{y}F_n'(u)\,\dd u\\
&\le-\frac1{4y_n}
\int_{y_n}^{y}(u-y_n)\,\dd u\\
&=-\frac{(y-y_n)^2}{8y_n}.
\end{aligned}
\end{equation}
Similarly, for \(y_n/2\le y\le y_n\), we have
\begin{equation}
\begin{aligned}
F_n(y)-F_n(y_n)&=-\int_y^{y_n}F_n'(u)\,\dd u\\
&\le-\frac1{4y_n}
\int_y^{y_n}(y_n-u)\,\dd u\\
&=-\frac{(y-y_n)^2}{8y_n}.
\end{aligned}
\end{equation}
Combining the two cases yields
\begin{equation}
F_n(y)-F_n(y_n)
\le-\frac{(y-y_n)^2}{8y_n},
\quad
\frac{y_n}{2}\le y\le2y_n.
\label{eq:app-saddle-Gaussian-tail}
\end{equation}
Then we apply this bound to the intermediate
tails outside the central window.
For fixed \(K>0\) and sufficiently
large \(n\), define
\begin{equation}
\mathcal I_{n,K}
:={}
\left[\frac{y_n}{2},
y_n-K\sqrt{y_n}\right]\cup
\left[y_n+K\sqrt{y_n},2y_n\right].
\end{equation}
Since \(y_n/2\ge Y\) for large \(n\),
Eq.~\eqref{eq:Haas_density} and
\(1/2\le\chi'(y)\le3/2\) give
\begin{equation}
y^nf_R(y)
\le CA_\varepsilon e^{F_n(y)},
\quad y\in\mathcal I_{n,K},
\end{equation}
where $
A_\varepsilon=
(1+\varepsilon)\sqrt{\frac32}.$
Applying
Eq.~\eqref{eq:app-saddle-Gaussian-tail},
we obtain
\begin{equation}\label{eq:app-saddle-intermediate}
\begin{aligned}
\frac{\displaystyle
\int_{\mathcal I_{n,K}}
y^nf_R(y)\,\dd y}
{C\sqrt{y_n}e^{F_n(y_n)}}&\le
\frac{A_\varepsilon}{\sqrt{y_n}}
\int_{\mathcal I_{n,K}}
e^{F_n(y)-F_n(y_n)}\,\dd y\\
&\le
\frac{A_\varepsilon}{\sqrt{y_n}}
\int_{\mathcal I_{n,K}}
e^{-(y-y_n)^2/(8y_n)}\,\dd y\\
&\le
2A_\varepsilon
\int_K^\infty e^{-z^2/8}\,\dd z,
\end{aligned}
\end{equation}
where the last inequality follows from
the substitution
\(z=(y-y_n)/\sqrt{y_n}\)
and the symmetry of the Gaussian
integrand.
Since $
\int_K^\infty e^{-z^2/8}\,\dd z
\longrightarrow0\ (K\to\infty),$
the normalized intermediate-tail
contribution can be made arbitrarily
small by choosing \(K\) sufficiently large. 

To complete the localization argument,
we show that the far-right and far-left
contributions are negligible relative
to the saddle scale. For \(y\ge2y_n\), the relation
\(\chi(y)/y\to1\) gives
\(\chi(y)/y\ge3/4\) for sufficiently
large \(n\).
Since \(n/y\le1/2\), we obtain
\begin{equation}
F_n'(y)
=\frac{n}{y}-\frac{\chi(y)}{y}
\le-\frac14.
\end{equation}
The Gaussian bound in
Eq.~\eqref{eq:app-saddle-Gaussian-tail}
at \(y=2y_n\) gives 
\begin{equation}\label{eq:bound_of_Fn_1}
F_n(2y_n)-F_n(y_n)
\le-\frac{y_n}{8}.
\end{equation}
For \(y\ge2y_n\), integrating
\(F_n'(u)\le-1/4\) over
\([2y_n,y]\) yields
\begin{equation}\label{eq:bound_of_Fn_2}
F_n(y)-F_n(2y_n)
=\int_{2y_n}^{y}F_n'(u)\,\dd u
\le-\frac{y-2y_n}{4}.
\end{equation}
Combining these two bounds in Eqs.~\eqref{eq:bound_of_Fn_1} and ~\eqref{eq:bound_of_Fn_2}, we obtain
\begin{equation}
F_n(y)-F_n(y_n)
\le-\frac{y_n}{8}
-\frac{y-2y_n}{4},
\quad y\ge2y_n.
\end{equation}
Using the density bound for \(y\ge Y\),
we find that
\begin{equation}\label{eq:far_right_vanish}
\begin{aligned}
\frac{\displaystyle
\int_{2y_n}^{\infty}
y^nf_R(y)\,\dd y}
{C\sqrt{y_n}e^{F_n(y_n)}}&\le
\frac{A_\varepsilon e^{-y_n/8}}
{\sqrt{y_n}}
\int_{2y_n}^{\infty}
e^{-(y-2y_n)/4}\,\dd y\\
&\quad=
4A_\varepsilon
\frac{e^{-y_n/8}}{\sqrt{y_n}}
\longrightarrow0,\qquad(n\to\infty).
\end{aligned}
\end{equation}
For the far-left region
\(Y\le y\le y_n/2\),
the monotonicity of \(F_n\) on
\([Y,y_n]\) gives
\(F_n(y)\le F_n(y_n/2)\).
Applying
Eq.~\eqref{eq:app-saddle-Gaussian-tail}
at \(y=y_n/2\), we obtain $
F_n(y)-F_n(y_n)
\le- y_n/32.$
Consequently, if $y\in[Y,y_n/2]$ we have 
\begin{equation}\label{eq:far_left_vanish}
\begin{aligned}
\frac{\displaystyle
\int_Y^{y_n/2}
y^nf_R(y)\,\dd y}
{C\sqrt{y_n}e^{F_n(y_n)}}
&\le
\frac{A_\varepsilon}{\sqrt{y_n}}
\left(\frac{y_n}{2}-Y\right)
e^{-y_n/32}\\
&\quad\le
\frac{A_\varepsilon}{2}
\sqrt{y_n}e^{-y_n/32}
\longrightarrow0,\quad (n\to\infty) .
\end{aligned}
\end{equation}
Thus, both far-tail contributions
vanish relative to the saddle scale according to Eq.~\eqref{eq:far_left_vanish} and Eq.~\eqref{eq:far_right_vanish}.

Finally, we combine the central and tail estimates
to establish the asymptotic equivalent
for the stationary moments.
For fixed \(\varepsilon\in(0,1)\) and
\(K>0\), consider the decomposition
\begin{equation}\label{eq:final_moment_decomposition}
\E R^n
=
I_{n,K}
+\int_{\mathcal I_{n,K}}
y^nf_R(y)\,\dd y
+\mathcal R_n,
\end{equation}
where \(I_{n,K}\) denotes the central-window
integral defined above and
\begin{equation}
\begin{aligned}
\mathcal R_n
&:={}\int_0^Y y^nf_R(y)\,\dd y+\int_Y^{y_n/2}
y^nf_R(y)\,\dd y\\&\qquad+\int_{2y_n}^{\infty}
y^nf_R(y)\,\dd y.
\end{aligned}
\end{equation}
This decomposition holds for sufficiently
large \(n\), when the central window
lies inside \([y_n/2,2y_n]\).

Note that the bounded-interval estimate in
Eq.~\eqref{eq:app-saddle-bounded},
together with the far-left and far-right
estimates in
Eqs.~\eqref{eq:far_left_vanish}
and~\eqref{eq:far_right_vanish},
gives
\begin{equation}\label{eq:vanish_Rn}
\frac{\mathcal R_n}
{C\sqrt{y_n}e^{F_n(y_n)}}
\longrightarrow0.
\end{equation}
Moreover, the central-window limit in
Eq.~\eqref{eq:limit_of_central}
yields
\begin{equation}
\frac{I_{n,K}}
{C\sqrt{y_n}e^{F_n(y_n)}}
\longrightarrow
\int_{-K}^{K}e^{-z^2/2}\,\dd z.
\end{equation}
For the intermediate-tail bound,
Eq.~\eqref{eq:app-saddle-intermediate}
further gives
\begin{equation}\label{eq:final_limit_central_window_tail}
0\le
\frac{\displaystyle
\int_{\mathcal I_{n,K}}
y^nf_R(y)\,\dd y}
{C\sqrt{y_n}e^{F_n(y_n)}}\le
2A_\varepsilon
\int_K^\infty e^{-z^2/8}\,\dd z,
\end{equation}
where
\(A_\varepsilon=(1+\varepsilon)\sqrt{3/2}\).

Since the integrand is nonnegative,
the full moment integral is bounded below
by its central-window contribution.
Taking the lower limit and using
Eq.~\eqref{eq:limit_of_central} and Eq.~\eqref{eq:final_moment_decomposition},
we obtain
\begin{equation}
\begin{aligned}
\liminf_{n\to\infty}
\frac{\E R^n}
{C\sqrt{y_n}e^{F_n(y_n)}}
&\ge
\lim_{n\to\infty}
\frac{I_{n,K}}
{C\sqrt{y_n}e^{F_n(y_n)}}\\
&\qquad=
\int_{-K}^{K}e^{-z^2/2}\,\dd z.
\end{aligned}
\end{equation}

For the upper limit, the same
decomposition shown in Eq.~\eqref{eq:final_moment_decomposition}, together with the
vanishing remainder \(\mathcal R_n\) shown in Eq.~\eqref{eq:vanish_Rn}
and the intermediate-tail bound in
Eq.~\eqref{eq:final_limit_central_window_tail},
gives
\begin{equation}
\begin{aligned}
\limsup_{n\to\infty}
\frac{\E R^n}
{C\sqrt{y_n}e^{F_n(y_n)}}
\le&
\int_{-K}^{K}e^{-z^2/2}\,\dd z\\
&+
2A_\varepsilon
\int_K^\infty e^{-z^2/8}\,\dd z.
\end{aligned}
\end{equation}
Combining these bounds, we obtain
\begin{equation}
\begin{aligned}
\int_{-K}^{K}e^{-z^2/2}\,\dd z&\le
\liminf_{n\to\infty}
\frac{\E R^n}
{C\sqrt{y_n}e^{F_n(y_n)}}\\
&\le
\limsup_{n\to\infty}
\frac{\E R^n}
{C\sqrt{y_n}e^{F_n(y_n)}}\\
&\le
\int_{-K}^{K}e^{-z^2/2}\,\dd z +
2A_\varepsilon
\int_K^\infty e^{-z^2/8}\,\dd z.
\end{aligned}
\end{equation}
For fixed \(\varepsilon\), the constant
\(A_\varepsilon\) is finite.
As \(K\to\infty\),
\begin{equation}
\int_{-K}^{K}e^{-z^2/2}\,\dd z
\longrightarrow\sqrt{2\pi},
\end{equation}
whereas $
2A_\varepsilon
\int_K^\infty e^{-z^2/8}\,\dd z
\longrightarrow0.$
Therefore, the lower and upper limits 
coincide, yielding
\begin{equation}
\lim_{n\to\infty}
\frac{\E R^n}
{C\sqrt{y_n}e^{F_n(y_n)}}
=\sqrt{2\pi}.
\end{equation}
Since
\(F_n(y_n)=n\log y_n-J(y_n)\),
we conclude that
\begin{equation}
\E R^n
\sim
C\sqrt{2\pi y_n}\,
y_n^n e^{-J(y_n)}.
\label{eq:moment_laplace}
\end{equation}
By matching this equivalent with
the exact stationary moments in
Eq.~\eqref{eq:stationary_moments_app}, we determine the prefactor in the density asymptotics.
In particular,
Eq.~\eqref{eq:moment_laplace} gives
\begin{equation}\label{limit_of_prefactor_C}
C=\lim_{n\to\infty}
\frac{n!e^{J(y_n)}}
{\Pi_n\sqrt{2\pi y_n}\,y_n^n}.
\end{equation}
Evaluating Eq.~\eqref{eq:Haas_density}
at \(y_n\) and eliminating \(C\)
using Eq.~\eqref{limit_of_prefactor_C}, we obtain
\begin{equation}
f_R(y_n)\sim
\frac{n!}{\Pi_n}
\frac{\sqrt{\chi'(y_n)}}
{\sqrt{2\pi y_n}\,y_n^n}.
\end{equation}
Applying Stirling's formula,
\(n!\sim\sqrt{2\pi n}(n/e)^n\) yields
\begin{equation}
f_R(y_n)
\sim\Pi_n^{-1}
\sqrt{\frac{n\chi'(y_n)}{y_n}}
\exp\left[
n\log\frac{n}{y_n}-n
\right].
\label{eq:density_at_saddle}
\end{equation}

\subsection{The moving-saddle equivalent}
We now transfer the density asymptotics
from the moment saddle \(y_n\) to a
general large threshold \(x\).
Choose \(n_x=\lfloor\chi(x)\rfloor\).
Since \(0\le\chi(x)-n_x<1\) and
\(g'(s)\to1\), the mean-value theorem gives
\begin{equation}
0\le x-y_{n_x}=g(\chi(x))-g(n_x)=O(1).
\label{eq:ynx_shift}
\end{equation}
Since \(\chi(x)\sim x\), we have
\(n_x=\lfloor\chi(x)\rfloor\to\infty\).
Moreover, \(\phi(n)\to1\) implies
\(y_{n_x}\sim n_x\sim x\to\infty\).
Thus, Eq.~\eqref{eq:Haas_density}
applies at both \(x\) and \(y_{n_x}\).
Writing the relative remainder as
\(\delta(y)=O(1/\chi(y))\), we obtain $
\delta(x)=O(x^{-1})\to0$ and $
\delta(y_{n_x})=O(n_x^{-1})\to0$.
Therefore,
\begin{equation}
\frac{f_R(x)}{f_R(y_{n_x})}
={}
\sqrt{\frac{\chi'(x)}
{\chi'(y_{n_x})}}
\exp\left[
-\int_{y_{n_x}}^x
\frac{\chi(u)}u\,\dd u
\right]
[1+o(1)].
\end{equation}
The bounded displacement also gives
\(\sqrt{\chi'(x)/\chi'(y_{n_x})}\to1\).
To transfer the exponential factor,
we use \(\chi(y_{n_x})=n_x\) and
\(n_x\le\chi(x)<n_x+1\).
Since \(\chi\) is increasing, $
0\le\chi(u)-n_x<1$ and
$y_{n_x}\le u\le x.$
Therefore, we write
\begin{equation}
\int_{y_{n_x}}^x
\frac{\chi(u)}{u}\,\dd u=n_x\log\frac{x}{y_{n_x}}
+\int_{y_{n_x}}^x
\frac{\chi(u)-n_x}{u}\,\dd u.
\end{equation}
The remainder satisfies
\begin{equation}
\begin{aligned}
0\le\int_{y_{n_x}}^x
\frac{\chi(u)-n_x}{u}\,\dd u\le\log\frac{x}{y_{n_x}}\le\frac{x-y_{n_x}}{y_{n_x}}
=O(x^{-1}),
\end{aligned}
\end{equation}
where the last estimate follows from
Eq.~\eqref{eq:ynx_shift} and
\(y_{n_x}\sim x\).
Consequently,
\begin{equation}
\int_{y_{n_x}}^x
\frac{\chi(u)}{u}\,\dd u
=n_x\log\frac{x}{y_{n_x}}+o(1).
\end{equation}
Combining this estimate with
\(\chi'(x)/\chi'(y_{n_x})\to1\)
and the vanishing relative errors
in Eq.~\eqref{eq:Haas_density}, we obtain
\begin{equation}\label{eq:ratio_of_threshold_saddle}
\frac{f_R(x)}{f_R(y_{n_x})}
\sim
\left(\frac{y_{n_x}}{x}\right)^{n_x}.
\end{equation}
Substituting
Eq.~\eqref{eq:density_at_saddle} into Eq.~\eqref{eq:ratio_of_threshold_saddle} gives
\begin{equation}
f_R(x)\sim\Pi_{n_x}^{-1}
\sqrt{\frac{n_x\chi'(x)}{x}}
\exp\left[n_x\log\frac{n_x}{x}-n_x\right].
\label{eq:density_moving_saddle_full}
\end{equation}
Since \(n_x/x\to1\) and
\(\chi'(x)\to1\), the square-root
factor tends to one, proving
Eq.~\eqref{eq:density-moving-saddle}.
Combining this density equivalent with
the MFPT relation in
Eq.~\eqref{eq:MFPT_Kx}, we obtain
\begin{equation}
\begin{aligned}
\E T_b
&\sim\frac{1}{rf_R(x)}\\
&\sim\frac{\Pi_{n_x}}{r}
\exp\left[
n_x-n_x\log\frac{n_x}{x}
\right]\\
&=\frac{e^x}{r}\Pi_{n_x}
\exp[-D(x,n_x)],
\end{aligned}
\end{equation}
where $D(x,n_x)=x-n_x+n_x\log(n_x/x)$, which establishes
Eq.~\eqref{eq:moving_saddle_MFPT}.

The rounding of the continuous saddle
\(\chi(x)\) to the integer index
\(n_x=\lfloor\chi(x)\rfloor\)
does not affect the displacement penalty
at the level of multiplicative asymptotics.
Indeed, since
\(\partial_nD(x,n)=\log(n/x)\)
and \(0\le\chi(x)-n_x<1\),
the mean-value theorem gives
\[
D(x,n_x)-D(x,\chi(x))
=(n_x-\chi(x))
\log\frac{\xi_x}{x},
\]
for some
\(\xi_x\in[n_x,\chi(x)]\).
Since \(n_x/x\to1\) and
\(\chi(x)/x\to1\), we have
\(\xi_x/x\to1\), and hence
\begin{equation}
D(x,n_x)-D(x,\chi(x))=o(1).
\label{eq:rounding_D}
\end{equation}
Here, we can conclude that $
e^{-D(x,n_x)}
\sim e^{-D(x,\chi(x))}.$
\subsection{Short-gap asymptotics and accuracy regimes}
Let us now impose the additional short-gap
power-law condition \(w(t)\sim ct^{\kappa-1}\) as \(t\downarrow0\),
where \(c>0\) and \(\kappa>0\).
Under this assumption, we determine the
relative accuracy of the fixed-index
product using the moving-saddle equivalent
established above.

To obtain the large-\(s\) equivalents, split
the Laplace integral for \(q(s)\) at any
small fixed \(\delta>0\). The portion over
\([\delta,\infty)\) is at most
\(e^{-\gamma s\delta}\), whereas on
\((0,\delta)\), comparison with
\(ct^{\kappa-1}\) and the substitution
\(v=\gamma st\) give the Gamma integral.
For the derivative, we have the exact identity
\begin{equation}
-q'(s)=\gamma\int_0^\infty
 te^{-\gamma st}w(t)\,\dd t.
\end{equation}
Its tail is at most
\(\gamma m e^{-\gamma s\delta}\), and the
same substitution gives \(\Gamma(\kappa+1)\)
in the leading term. Hence
\begin{equation}
q(s)\sim a s^{-\kappa},
\quad q'(s)\sim-\kappa a s^{-\kappa-1}.
\label{eq:q_qprime}
\end{equation}
Since \(\chi(x)\sim x\) and
\(x-\chi(x)=xq(\chi(x))\), we obtain
\begin{equation}
1-\frac{\chi(x)}x
=q(\chi(x))\sim a x^{-\kappa},
\label{eq:chi_shift_rel}
\end{equation}
and hence
\begin{equation}
x-\chi(x)\sim a x^{1-\kappa}.
\label{eq:chi_shift}
\end{equation}

To compare with the deterministic product,
put \(m_x=\lfloor x\rfloor\). The
moving-saddle equivalent of Eq.~\eqref{eq:moving_saddle_MFPT} gives
\begin{equation}
\log\frac{\E T_b}{T_{\mathrm{prod}}(b)}
=\log\frac{\Pi_{n_x}}{\Pi_{m_x}}-D(x,n_x)+o(1).
\label{eq:log_ratio_master}
\end{equation}
Writing \(\epsilon_x=1-n_x/x\), the floor
convention yields
\begin{equation}
\epsilon_x=q(\chi(x))+O(x^{-1}).
\label{eq:epsilon_round}
\end{equation}
For \(0<\kappa<1\), this implies
\(\epsilon_x\sim a x^{-\kappa}\). More generally,
\begin{equation}
D(x,n_x)
=x[\epsilon_x+(1-\epsilon_x)\log(1-\epsilon_x)]
=\frac{x\epsilon_x^2}{2}[1+O(|\epsilon_x|)].
\label{eq:D_expand}
\end{equation}
Thus \(D(x,n_x)\sim(a^2/2)x^{1-2\kappa}\)
for \(0<\kappa\le1/2\), whereas
\(D(x,n_x)\to0\) for every \(\kappa>1/2\).

For \(0<\kappa<1\), the saddle
displacement and the floor convention give
\begin{equation}
m_x-n_x
=x-\chi(x)+O(1)
\sim ax^{1-\kappa}.
\end{equation}
Since \(q\) is decreasing, for
\(n_x<j\le m_x\) we have
\begin{equation}\label{q_endpoints}
q(m_x)\le q(j)\le q(n_x+1).
\end{equation}
Both endpoints in Eq.~\eqref{q_endpoints} are asymptotic to
\(q(x)\sim ax^{-\kappa}\), because
\(n_x/x\to1\) and \(m_x/x\to1\).
Thus,
\(q(j)\sim q(x)\) uniformly
over this interval.
By the definition of \(\Pi_n\), $
\log\frac{\Pi_{n_x}}{\Pi_{m_x}}
=-\sum_{j=n_x+1}^{m_x}
\log[1-q(j)].$
Since \(q(j)\to0\) uniformly on the
summation interval, the expansion
\(-\log(1-u)=u+O(u^2)\)
holds uniformly for \(u=q(j)\).
Hence,
\begin{equation}
\log\frac{\Pi_{n_x}}{\Pi_{m_x}}=\sum_{j=n_x+1}^{m_x}q(j)+
O\left(
\sum_{j=n_x+1}^{m_x}q(j)^2
\right).
\end{equation}
Using the uniform equivalent for \(q(j)\)
and \(m_x-n_x\sim ax^{1-\kappa}\),
we obtain
\begin{align}
\sum_{j=n_x+1}^{m_x}q(j)
&\sim(m_x-n_x)q(x)
\sim a^2x^{1-2\kappa},
\label{eq:sum_q}\\
\sum_{j=n_x+1}^{m_x}q(j)^2
&=O(x^{1-3\kappa})
=o(x^{1-2\kappa}).
\label{eq:sum_q2}
\end{align}
Hence, for \(0<\kappa<1\),
\begin{equation}
\log\frac{\Pi_{n_x}}{\Pi_{m_x}}
\sim a^2x^{1-2\kappa}.
\end{equation}

When \(\kappa\ge1\), the saddle
displacement satisfies
\(x-\chi(x)=O(1)\), so
\(m_x-n_x=O(1)\).
Since \(q(j)=O(x^{-\kappa})\)
for the intervening indices, $
\log\frac{\Pi_{n_x}}{\Pi_{m_x}}
=O(x^{-\kappa})=o(1).$
Combining these product estimates with
the displacement expansion in
Eq.~\eqref{eq:D_expand}
and the logarithmic decomposition in
Eq.~\eqref{eq:log_ratio_master}
establishes
Eqs.~\eqref{eq:regime_super}--\eqref{eq:regime_sub}.
At \(\kappa=1/2\), the product gain
tends to \(a^2\), while the displacement
penalty tends to \(a^2/2\), yielding
the finite correction \(e^{a^2/2}\).
For \(0<\kappa<1/2\), their difference
grows as \((a^2/2)x^{1-2\kappa}\)
on the logarithmic scale.

\section{Refined critical asymptotics}
\label{app:refined}

The critical regime admits a refined
MFPT equivalent when the short-time
waiting-time density has a higher-order
expansion. We assume
\begin{equation}
w(t)=ct^{-1/2}+d_0+O(t^{1/2}),
\qquad t\downarrow0,
\label{eq:critical_w}
\end{equation}
where \(c>0\).
This assumption determines the
subleading terms in the renewal product
and, together with the critical
correction in
Eq.~\eqref{eq:regime_critical},
yields a full multiplicative
MFPT equivalent. Then we have
\begin{equation}\label{eq:critical_q_app}
q(s)=a s^{-1/2}+\eta s^{-1}+O(s^{-3/2}),
\end{equation}
where $a=\frac{c\sqrt{\pi}}{\sqrt{\gamma}}$ and
$\eta=\frac{d_0}{\gamma}$. Since \(q(n)=O(n^{-1/2})\), expanding
the logarithm gives
\begin{equation}
\begin{aligned}
\log[1-q(n)]
&=-q(n)-\frac12q(n)^2+O(q(n)^3)
\\
&=-a n^{-1/2}
-\left(\eta+\frac{a^2}{2}\right)n^{-1}
+O(n^{-3/2}).
\end{aligned}
\label{eq:critical_log_factor_app}
\end{equation}
The \(O(n^{-3/2})\) remainder is absolutely
summable. Note that $
 \sum_{j=1}^n j^{-1/2}=2\sqrt n+C_{1/2}+o(1),$
and $\sum_{j=1}^n j^{-1}=\log n+\gamma_{\rm E}+o(1),$
where \(C_{1/2}\) is finite and
\(\gamma_{\rm E}\) is Euler's constant.
Summing Eq.~\eqref{eq:critical_log_factor_app},
with the finitely many initial indices absorbed
into a constant, therefore gives
\begin{equation}
 \log\Pi_n=-2a\sqrt n
 -(\eta+a^2/2)\log n+C_{\rm prod}+o(1),
\end{equation}
for a finite real constant \(C_{\rm prod}\).
Exponentiation shows that the positive constant
\begin{equation}
B_{\mathrm{prod}}
:=e^{C_{\rm prod}}=
\lim_{n\to\infty} 
e^{2a\sqrt{n}} 
n^{\eta+a^2/2}\Pi_n
\in(0,\infty)
\label{eq:Bprod}
\end{equation}
is well defined, with
\begin{equation}
\Pi_n
\sim
B_{\mathrm{prod}}
 e^{-2a\sqrt n}
 n^{-(\eta+a^2/2)}.
\label{eq:Pi_critical}
\end{equation}
Because \(\lfloor x\rfloor=x+O(1)\),
Eq.~\eqref{eq:Pi_critical} also gives
\(\Pi_{\lfloor x\rfloor}\sim B_{\mathrm{prod}}
e^{-2a\sqrt{x}}x^{-(\eta+a^2/2)}\).
Combining this equivalent with the
critical MFPT ratio in Eq.~\eqref{eq:regime_critical}
yields
\begin{equation}
\E T_b
\sim
\frac{B_{\mathrm{prod}}e^{a^2/2}}{r}
 e^{x-2a\sqrt{x}}
 x^{-(\eta+a^2/2)}.
\label{eq:critical_full_app}
\end{equation}

\section{Full first-passage transform and limiting distribution}
\label{app:laplace}
Let us return to the general waiting-time assumptions: the
intervals are i.i.d.\ and strictly positive, with
\(m=\E W\in(0,\infty)\), and are independent
of the exponential marks. The distribution of
\(W\) need not have a density.

For \(\theta\ge0\), we write
\(\widehat w(\theta)=\E e^{-\theta W}\)
for the Laplace transform of the waiting-time law.
This notation is related to the scaled transform in the main
text by \(q(s)=\widehat w(\gamma s)\). Define
\begin{equation}
\mathcal F_{\theta}(y)
:=
\sum_{n=0}^{\infty}\frac{y^n}{n!}
\prod_{j=0}^{n-1}\bigl[1-\widehat w(\theta+j\gamma)\bigr],
\label{eq:Ftheta_def}
\end{equation}
where the empty product is one. For \(0\le y<x\), we define the first-passage
Laplace transform by
\begin{equation}
L_\theta(y)
:=\E\!\left[e^{-\theta T_b^{(y/\lambda)}}\right],
\qquad 0\le y<x.
\end{equation}
Here \(T_b^{(y/\lambda)}\) corresponds to
the initial signal level \(X(0)=y/\lambda\),
or equivalently \(R_0=y\), at the beginning
of a fresh renewal interval.
Then, we define
\begin{equation}
H_\theta(y)
:=e^{y-x}+\int_y^x e^{y-z}L_\theta(z)\,\dd z.
\label{eq:Htheta}
\end{equation}
To obtain the first-passage transform, we condition
on the first renewal interval and exponential mark.
Given \(Qy=a<x\), crossing at the first arrival
occurs with probability \(e^{-(x-a)}\). Otherwise,
the process restarts at \(a+E\), with future innovations independent of the first pair.
The exponential density of the mark therefore gives
\begin{equation}\label{eq:L_theta}
L_\theta(y)=\E\Bigl[e^{-\theta W}\Bigl\{
e^{-(x-Qy)}+\int_0^{x-Qy}e^{-v}L_\theta(Qy+v)\,\dd v
\Bigr\}\Bigr].    
\end{equation}
The expression in braces of Eq.~\eqref{eq:L_theta} equals \(H_\theta(Qy)\),
so the transform and auxiliary function satisfy
\begin{align}
L_\theta(y)
&=\E\!\left[e^{-\theta W}H_\theta(Qy)\right],
\label{eq:Laplace_backward}
\\
H_\theta'(y)&=H_\theta(y)-L_\theta(y),
\quad H_\theta(x)=1.
\label{eq:Htheta_ode}
\end{align}
The coefficients of \(\mathcal F_\theta\) obey
\(c_0=1\) and
\(c_{n+1}=c_n[1-\widehat w(\theta+n\gamma)]\).
Their absolute values are at most one, making
\(\mathcal F_\theta\) entire. Comparing successive
coefficients after termwise differentiation yields
\begin{equation}
\mathcal F_\theta'(y)=\mathcal F_\theta(y)
-\E\!\left[e^{-\theta W}\mathcal F_\theta(Qy)\right].
\end{equation}
Consequently, the candidate
\(\widetilde H_\theta(y)=\mathcal F_\theta(y)/
\mathcal F_\theta(x)\) satisfies the differential
relation in Eq.~\eqref{eq:Htheta_ode} and the
boundary condition \(\widetilde H_\theta(x)=1\).
Defining \(\widetilde L_\theta(y)=
\E[e^{-\theta W}\widetilde H_\theta(Qy)]\),
the differential relation also gives the integral
representation in Eq.~\eqref{eq:Htheta}.
For \(\theta>0\), the original first-step operator
on bounded transforms is a contraction with factor
\(\widehat w(\theta)<1\), so its bounded solution
is unique. At \(\theta=0\), both
\(L_0\equiv1\) and \(\mathcal F_0\equiv1\).
Thus, for every \(\theta\ge0\),
\begin{equation}
H_\theta(y)=\frac{\mathcal F_\theta(y)}{\mathcal F_\theta(x)},
\end{equation}
with \(\mathcal F_\theta\) as defined in
Eq.~\eqref{eq:Ftheta_def}. Evaluating
Eq.~\eqref{eq:Laplace_backward} at \(y=0\) yields
\begin{equation}
L_\theta(0)=\frac{\widehat w(\theta)}{\mathcal F_\theta(x)}.
\label{eq:laplace_exact}
\end{equation}
This is the exact Laplace transform of the
first-passage time. To justify differentiation at
\(\theta=0\), let
\(c_n(\theta)=\prod_{j=0}^{n-1}[1-\widehat w(\theta+j\gamma)]\).
Since \(0\le c_n(\theta)\le1\) and
\(0\le-\widehat w'(\theta)\le m\), the product rule gives
\(0\le c_n'(\theta)\le nm\) for \(\theta\ge0\).
For fixed \(x\), the derivative series is therefore
uniformly bounded by
\(\sum_{n\ge1}(x^n/n!)nm=mxe^x\).
Termwise differentiation is justified, and
\(\partial_\theta\mathcal F_\theta(x)|_{\theta=0}
=mV_0(x)\), while \(\widehat w'(0)=-m\).
Differentiating the quotient at zero therefore
gives \(-\left.\frac{\partial L_\theta(0)}{\partial\theta}\right|_{\theta=0}
=m[1+V_0(x)]\), in agreement with
Eq.~\eqref{eq:exact_MFPT}.

To obtain the high-threshold limit, we write
\begin{equation}\label{eq:relation_of_F_And_V_theta}
\mathcal F_\theta(x)
=1+[1-\widehat w(\theta)]V_\theta(x),
\end{equation}
with $
V_\theta(x)
:=\sum_{n\ge1}\frac{x^n}{n!}
\prod_{j=1}^{n-1}[1-\widehat w(\theta+j\gamma)].$
Comparing \(V_\theta\) with \(V_0\), monotonicity of \(\widehat w\) together with
\(\log(1+z)\le z\) and
\(1-e^{-\theta W}\le\theta W\), yields
\begin{equation}
1\le\frac{V_\theta(x)}{V_0(x)}
\le
\exp\!\left[
\frac{\theta}{\gamma[1-\widehat w(\gamma)]}
\right].
\label{eq:Vtheta_bound}
\end{equation}
Indeed, for \(j\ge1\), monotonicity of
\(\widehat w\) and the inequality
\(1-e^{-\theta W}\le\theta W\) imply
\begin{equation}
0\le
\log\frac{1-\widehat w(\theta+j\gamma)}
{1-\widehat w(j\gamma)}
\le
\frac{\theta\E[We^{-j\gamma W}]}
{1-\widehat w(j\gamma)}.
\end{equation}
Moreover,
\begin{equation}
\begin{aligned}
\sum_{j\ge1}
\frac{\E[We^{-j\gamma W}]}
{1-\widehat w(j\gamma)}
&\le
\frac{1}{1-\widehat w(\gamma)}
\E\left[\frac{W}{e^{\gamma W}-1}\right]\\
&\le
\frac{1}{\gamma[1-\widehat w(\gamma)]}.
\end{aligned}
\end{equation}
Applying these bounds term by term in
\(V_\theta(x)\) proves Eq.~\eqref{eq:Vtheta_bound}.
Set \(M_x=\E T_b=m[1+V_0(x)]\). Since \(V_0(x)\ge x\), we have \(M_x\to\infty\) as \(x\to\infty\).
For fixed \(z>0\), let
\(\theta_x=z/M_x\), so that
\(\theta_x\to0\). Thus, the uniform comparison in
Eq.~\eqref{eq:Vtheta_bound} gives
\begin{equation}
1\le
\frac{V_{\theta_x}(x)}{V_0(x)}
\le
\exp\left[
\frac{\theta_x}
{\gamma[1-\widehat w(\gamma)]}
\right]
\longrightarrow1,\quad (x\to\infty) .
\end{equation}
Moreover, since \(\E W=m<\infty\),
dominated convergence yields
\begin{equation}
\frac{1-\widehat w(\theta)}
{\theta}
=
\E\left[
\frac{1-e^{-\theta W}}{\theta}
\right]
\longrightarrow m
\qquad(\theta\downarrow0).
\end{equation}
Hence, \(1-\widehat w(\theta_x)\sim
m\theta_x\). Using
\(\mathcal F_\theta(x)
=1+[1-\widehat w(\theta)]V_\theta(x)\) derived from Eq.~\eqref{eq:relation_of_F_And_V_theta},
we obtain
\begin{equation}\label{eq:F_theta_x}
\mathcal F_{\theta_x}(x)-1=
\frac{1-\widehat w(\theta_x)}
{m\theta_x}
\frac{V_{\theta_x}(x)}{V_0(x)}
\,m\theta_xV_0(x).
\end{equation}
Note that the first two factors tend to one in Eq.~\eqref{eq:F_theta_x}, while $
m\theta_xV_0(x)=z\frac{V_0(x)}{1+V_0(x)}\longrightarrow z.$
Therefore,
\begin{equation}
\mathcal F_{z/M_x}(x)
\longrightarrow1+z.
\end{equation}
Since \(\theta_x\to0\), we also have
\(\widehat w(\theta_x)\to1\). Then, the exact first-passage transform
in Eq.~\eqref{eq:laplace_exact} yields
\begin{equation}
\E e^{-zT_b/M_x}=\frac{\widehat w(z/M_x)}
{\mathcal F_{z/M_x}(x)}\longrightarrow\frac1{1+z},
\quad z>0.
\end{equation}
Note that the same limit holds at \(z=0\).
Since \(1/(1+z)\) is the Laplace
transform of \(\Exp(1)\), the
continuity theorem for Laplace
transforms establishes
Eq.~\eqref{eq:exp_limit}.

\section{Numerical evaluation and event-driven simulation}
\label{app:numerics}

The figures illustrated in our paper use two distinct numerical procedures.
Fig.~\ref{fig:model}(a) shows a selected
event-driven trajectory. Figs.~\ref{fig:model}(b),
\ref{fig:crossover}, \ref{fig:regimes}, and
\ref{fig:moving_saddle_accuracy}, as well as Figs.~S1 and ~S2
in the Supplemental Material \cite{SupplementalMaterial}, use deterministic
evaluations of the auxiliary-index weights,
exact MFPT series, or moving-saddle expressions. Fig.~\ref{fig:exact}
compares Eq.~\eqref{eq:exact_MFPT} with
independent event-driven means for Gamma and
Weibull waiting times. 

\subsection{Benchmark renewal families}
For Gamma renewal intervals of unit mean,
\begin{equation}
W\sim\mathrm{Gamma}(\kappa,\text{rate}=\kappa),
\end{equation}
the scaled waiting-time Laplace transform is
\begin{equation}
q(s)=\left(\frac{\kappa}{\kappa+\gamma s}\right)^\kappa.
\end{equation}
Throughout the numerical comparisons we set
\(\lambda=r=\gamma=1\) and choose
\(\kappa=0.75,0.50,0.25\) to represent
the weak, critical, and strong displacement regimes.

To separate the short-time exponent from the
full waiting-time law, we also consider a
unit-mean Weibull distribution with density
\begin{equation}
w(t)
=\frac{\kappa}{\beta}\left(\frac{t}{\beta}\right)^{\kappa-1}
\exp[-(t/\beta)^\kappa],
\quad
\beta=\Gamma(1+1/\kappa)^{-1}.
\end{equation}
For \(0<\kappa<1\), the scaled waiting-time
Laplace transform admits the convergent expansion
\begin{equation}
q(s)=\frac{\kappa}{(\gamma\beta s)^\kappa}
\sum_{m=0}^{\infty}
\frac{(-1)^m\Gamma[\kappa(m+1)]}
{m!(\gamma\beta s)^{\kappa m}}.
\end{equation}
The figures use numerical evaluation of
$
q(s)=\E[e^{-\gamma sW}]\\
=\int_0^\infty
e^{-u-\gamma s\beta u^{1/\kappa}}\,\dd u,
$
which is valid for all \(\kappa>0\), rather than
truncating the series.
The integral was evaluated by Gauss--Legendre
quadrature with a scale-adapted variable and
384 nodes, and checked against independent
adaptive quadrature.

For the cross-family comparisons in
Fig.~\ref{fig:moving_saddle_accuracy} and 
Fig.~S1 in the Supplemental Material \cite{SupplementalMaterial}, we also use the mean-one generalized-Gamma
family
\begin{equation}
\begin{aligned}
 w_{\kappa,p}(t)
 =\frac{p\,t^{\kappa-1}}
 {\beta^\kappa\Gamma(\kappa/p)}
 \exp[-(t/\beta)^p], \ \beta=\frac{\Gamma(\kappa/p)}
{\Gamma((\kappa+1)/p)}.
\end{aligned}
\label{eq:GG_density}
\end{equation}
The choices $p=1$ and $p=\kappa$ recover the Gamma and
Weibull distributions above while the additional benchmark uses
$p=2$. All three share the short-time exponent $\kappa$
within each panel, with family-dependent Laplace coefficient
\begin{equation}
 a_{\kappa,p}
 =\frac{p\,\Gamma(\kappa)}
 {\gamma^{\kappa}\beta^{\kappa}\Gamma(\kappa/p)}.
 \label{eq:GG_coefficient}
\end{equation}
Numerical calculations use the full Laplace transforms
of the respective waiting-time distributions.
For the
generalized-Gamma family, the substitution
\(u=(t/\beta)^p\) gives the integral
\begin{equation}
q_{\kappa,p}(s)=\frac{1}{\Gamma(\kappa/p)}
\int_0^\infty u^{\kappa/p-1}
\exp[-u-\gamma s\beta u^{1/p}]\,\dd u.
\label{eq:GG_Laplace_integral}
\end{equation}
If the integration is truncated at \(u=U\), its omitted
nonnegative tail is at most
\(\Gamma(\kappa/p,U)/\Gamma(\kappa/p)\),
where the numerator is the upper incomplete Gamma function.
This bound provides a direct check on the truncation error,
independently of the quadrature method.
The coefficients in Eq.~\eqref{eq:GG_coefficient} enter only
the asymptotic normalization in Fig.~S1 of the Supplemental Material  \cite{SupplementalMaterial}.

\subsection{Stable exact-series evaluation}

To avoid overflow and loss of precision for
large \(x\), we evaluate
Eq.~\eqref{eq:exact_MFPT} in logarithmic form.
For \(x>0\) and \(n\ge1\), define
\begin{equation}
\ell_n=n\log x-\log\Gamma(n+1)+\log\Pi_{n-1},
\quad \ell_0=0.
\end{equation}
We sum the positive terms using log-sum-exp
and update the product factors through
\begin{equation}
\log\Pi_n=\log\Pi_{n-1}+\log[1-q(n)].
\end{equation}
The summation window is initially centered
near \(n\approx\chi(x)\) and enlarged
until the omitted contributions are controlled.
A deterministic right-tail bound follows from
Eq.~\eqref{eq:summand_ratio} whenever \(N+2>x\),
\begin{equation}
\sum_{n=N+1}^{\infty}C_n(x)
\le\frac{C_{N+1}(x)}{1-x/(N+2)},
\end{equation}
and the left tail can be evaluated directly by
retaining terms down to \(n=0\).
These bounds allow the omitted contributions
to be controlled relative to the retained
positive sum at a prescribed tolerance.

Note that the continuous saddle satisfies
\(\chi/[1-q(\chi)]=x\), whereas the
maximizing discrete index is
\begin{equation}
n_*(x):=\min\operatorname*{arg\,max}_{n\ge0}C_n(x).
\end{equation}
Thus \(n_*(x)\) is a lattice index
(with ties resolved by the smallest index), while \(\chi(x)\)
is defined by a continuous balance equation.
Fig.~\ref{fig:crossover}(a) uses the exact
normalized summands and Fig.~\ref{fig:crossover}(b)
shows the displacement of the continuous saddle.
Products and series are computed in logarithmic
form and checked against numerical tail bounds.
For Fig.~\ref{fig:moving_saddle_accuracy}, the relative
error is evaluated from the log-domain identity
\begin{equation}
\begin{aligned}
\varepsilon_{\mathrm{sad}}(x)
&=\operatorname{expm1}\!\left(
\log M_{\mathrm{sad}}-\log M_x\right),\\
\log M_{\mathrm{sad}}
&=x-\log r+\log\Pi_{n_x}-D(x,n_x),
\end{aligned}
\end{equation}
with $n_x=\lfloor\chi(x)\rfloor$, avoiding direct
evaluation of exponentially large MFPTs.

\subsection{Event-driven simulation}

For each trajectory, we start from
\(X=t=0\). At each arrival, we draw a new
waiting interval \(W\) and an independent
exponential mark \(A\), then update
\begin{equation}
X\leftarrow Xe^{-\gamma W}+A,
\qquad
t\leftarrow t+W.
\end{equation}
Since \(X\) decays between arrivals, threshold
crossings can occur only at arrival times.
The first update for which \(X>b\) gives
the first-passage time \(T_b\).
For Fig.~\ref{fig:exact}, we record the first
crossing times at \(x=\lambda b=2,\ldots,7\)
along each trajectory and continue until
the largest threshold \(b=7/\lambda\) is crossed.

For each combination of waiting-time family
(Gamma or Weibull) and shape parameter
\(\kappa\in\{0.25,0.50,0.75\}\), we simulate
\(M=22{,}000\) independent trajectories.
The simulated means are compared with the
exact MFPT series.
Error bars indicate pointwise approximate
95\% confidence intervals,
\(\overline T_b\pm1.96\,s_x/\sqrt{M}\),
where \(s_x\) is the sample standard deviation
of the crossing times at threshold \(x\).
Estimates at different thresholds within
the same ensemble are correlated because
they are obtained from the same trajectories.
\bibliography{ref,supplemental_entry}

\end{document}


\title{Supplemental Material for
``Exact mean first-passage time and a Poisson-smoothing crossover in renewal shot noise''}
\author{Yuxin Wang}
\author{Jiajun Zhang}
\affiliation{School of Mathematics, Sun Yat-sen University, Guangzhou, China}
\date{\today}
\maketitle

\begin{figure}[h]
\centering
\includegraphics[width=1\linewidth]{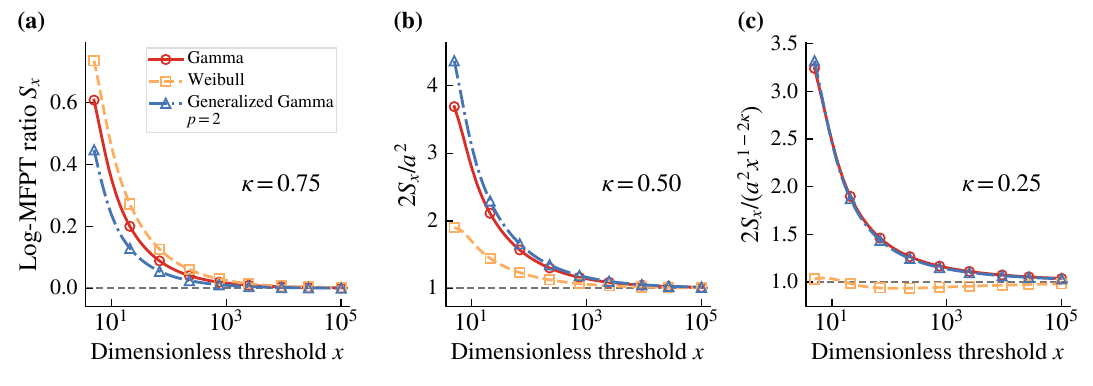}
\caption{Three accuracy regimes of the fixed-index MFPT approximation
across renewal waiting-time distributions.
The logarithmic discrepancy between the exact MFPT and the
fixed-index approximation is evaluated for three mean-one
waiting-time distributions: Gamma, Weibull, and generalized Gamma
with $p=2$. The generalized-Gamma density has shape
parameters $\kappa,p$, with $p=1$ and $p=\kappa$ recovering Gamma
and Weibull waiting times, respectively.
Red solid lines with open circles, orange dashed lines with open
squares, and blue dash-dotted lines with open triangles denote
Gamma, Weibull, and generalized Gamma ($p=2$), respectively.
Symbols mark selected deterministic evaluation points; gray dotted
lines indicate the predicted asymptotic limits.
Parameters are $\lambda=\gamma=r=1$, with exponential marks. These distributions share the same short-time exponent
$\kappa$ within each panel.
(a) For $\kappa=0.75$, the logarithmic ratio
$S_x=\log[M_x/T_{\mathrm{prod}}(b)]$ approaches zero.
(b) At $\kappa=0.50$, the normalized correction $2S_x/a^2$
approaches unity.
(c) For $\kappa=0.25$, the scaled logarithmic discrepancy
$2S_x/(a^2x^{1-2\kappa})$ approaches unity, with distinct
finite-threshold corrections across the three distributions.
Here $M_x=\mathbb{E}T_b$ is computed from the exact MFPT series,
$T_{\mathrm{prod}}(b)=e^x\Pi_{\lfloor x\rfloor}/r$,
$a=p\Gamma(\kappa)/[\gamma^\kappa\beta^\kappa\Gamma(\kappa/p)]$,
and $\beta=\Gamma(\kappa/p)/\Gamma((\kappa+1)/p)$ sets the mean
waiting time to unity. }
\label{fig:sm_fixed_index}
\end{figure}

\begin{figure}[p]
\centering
\includegraphics[
    width=0.98\textwidth,
    height=0.61\textheight,
    keepaspectratio
]{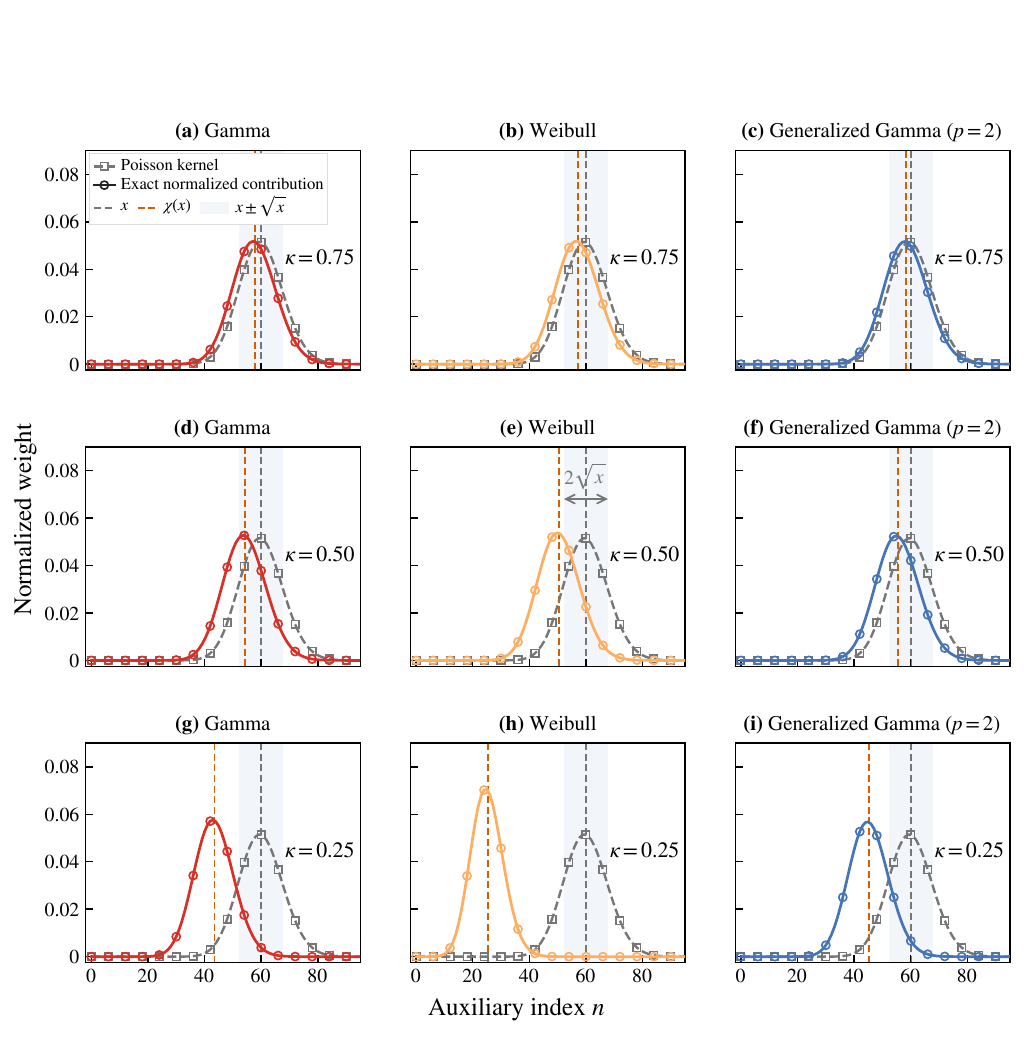}
\caption{Renewal-dependent displacement of the dominant
MFPT-series contributions across waiting-time distributions.
The auxiliary Poisson probabilities
$\mathbb{P}(N_x=n)$ (gray dashed lines, squares)
are compared with the normalized exact MFPT-series
contributions $\widetilde C_n(x)$ (colored solid lines,
circles) at $x=60$.
Columns correspond to Gamma (red), Weibull (orange),
and generalized-Gamma ($p=2$, blue) waiting times,
and rows to $\kappa=0.75$, $0.50$, and $0.25$.
The shaded bands span the one-standard-deviation
interval $[x-\sqrt{x},x+\sqrt{x}]$ around the
Poisson mean, with the width $2\sqrt{x}$ indicated
in panel (e).
Vertical dotted lines mark the Poisson center $x$
and the continuous saddle $\chi(x)$.
The renewal-dependent weighting shifts the dominant
MFPT contributions toward smaller indices, with the
displacement and profile shape varying across
waiting-time distributions and $\kappa$.
Symbols denote selected integer indices, and the
contributions are normalized by the full series.
All waiting-time distributions have unit mean,
with $\lambda=\gamma=r=1$ and exponential
marks.}
\label{fig:S2_mechanism}
\end{figure}

This Supplemental Material contains two deterministic comparisons supporting
the accuracy classification and its auxiliary-index interpretation.
Figure~\ref{fig:sm_fixed_index} compares fixed-index accuracy across renewal
families; Fig.~\ref{fig:S2_mechanism} resolves the corresponding index profiles.
Throughout, \(x=\lambda b\), \(q(s)=\E e^{-\gamma sW}\),
\(\Pi_0=1\), and \(\Pi_n=\prod_{j=1}^n[1-q(j)]\).
For \(n\ge1\), let \(C_n(x)=x^n\Pi_{n-1}/n!\), with \(C_0(x)=1\),
and \(\widetilde C_n(x)=C_n(x)/\sum_{j\ge0}C_j(x)\).
The continuous saddle solves \(\chi/[1-q(\chi)]=x\).
The bands in Fig.~\ref{fig:S2_mechanism} describe the standard deviation of the auxiliary Poisson kernel.

\clearpage
